\documentclass[sort&compress,12pt]{elsarticle}

\usepackage{amssymb}
\usepackage[english]{babel}
\usepackage{graphics}
\usepackage{axodraw}
\usepackage{rotating}
\usepackage{url}
\usepackage{xspace}
\usepackage{multirow}
\usepackage{amsmath}
\usepackage{comment}
\usepackage{tcolorbox}      
\usepackage{stackengine}
\usepackage{scalerel}
\usepackage{xcolor}
\usepackage{fancyvrb}   
\usepackage{marvosym}
\usepackage{hyperref}
\usepackage{relsize}   

\journal{Computer Physics Communications}
\makeatletter
  \def\ps@pprintTitle{%
     \let\@oddhead\@empty
     \let\@evenhead\@empty
     \def\@oddfoot{\footnotesize\itshape
     \hfill\today}%
     \let\@evenfoot\@oddfoot
}
\makeatother 

\def\ie{i.e.\ }
\def\eg{e.g.\ }

\newcommand{\RecolaTwo}{{\Huge \boldmath{\bf \sc R E C O L A 2}}}
\newcommand{\recola}{{\sc Recola}}
\newcommand{\recolatwo}{{\sc Recola2}}

\newcommand{\pole}{{\sc Pole}}
\newcommand{\openloops}{{\sc OpenLoops}}
\newcommand{\collier}{{\sc Collier}}
\newcommand{\CMake}{{\sc CMake}}
\newcommand{\Reptil}{{\sc REPT1L}}
\newcommand{\Feynrules}{{\sc Feynrules}}
\newcommand{\QGS}{{\tt{QGS}}}
\newcommand{\QGraf}{{\tt{QGRAF}}}
\newcommand{\GraphShot}{{\tt{GraphShot}}}
\newcommand{\FeynArts}{{\sc FeynArts}}
\newcommand{\FormCalc}{{\sc FormCalc}}
\newcommand{\FORM}{{\sc FORM}}
\newcommand{\Fortran}{{\sc Fortran}}
\newcommand{\FortranNinety}{{\sc Fortran95}}
\newcommand{\UFO}{{\sc UFO}}
\newcommand{\Rplus}{\protect\hspace{-.1em}\protect\raisebox{.35ex}{\smaller{\smaller\textbf{+}}}}
\newcommand{\Cpp}{\mbox{C\Rplus\Rplus}\xspace}
\newcommand{\CC}{\mbox{C}\xspace}
\newcommand{\Python}{{\sc Python}}
\newcommand{\PythonTwo}{{\sc Python 2.7}}
\newcommand{\PythonThree}{{\sc Python 3.x}}
\newcommand{\GeV}{\unskip\,\mathrm{GeV}}
\def\mathswitchr#1{\relax\ifmmode{\mathrm{#1}}\else$\mathrm{#1}$\fi}
\newcommand{\cmm}[1]{\ensuremath{#1}\ifmmode\else{}\fi}
\newcommand{\nmc}[2]{\newcommand{#1}{\cmm{#2}}}

\newcommand{\PW}{\mathswitchr W}

\newcommand{\PZ}{\mathswitchr Z}

\newcommand{\Pg}{\mathswitchr g}

\nmc{\PHl}{H_\mathrm{l}}
\nmc{\PHh}{H_\mathrm{h}}
\nmc{\PHa}{H_\mathrm{a}}
\nmc{\Hl}{\PHl}
\nmc{\Hh}{\PHh}
\nmc{\Ha}{\PHa}
\nmc{\Hpm}{\PHpm}
\nmc{\PG}{G_0}
\nmc{\GZ}{\PG}
\nmc{\PHpm}{H^\pm}
\nmc{\PGpm}{G^\pm}

\newcommand{\Phione}{\mathswitchr\Phi_1}
\newcommand{\Phitwo}{\mathswitchr\Phi_2}
\newcommand{\m}{\texttt{m}}

\newcommand{\ml}{\texttt{ml}}
\newcommand{\gl}{\texttt{gl}}
\newcommand{\mh}{\texttt{mh}}
\newcommand{\gh}{\texttt{gh}}
\nmc{\g}{g}
\nmc{\gy}{{g^\prime}}
\nmc{\sw}{s_\mathrm{w}}
\nmc{\cw}{c_\mathrm{w}}
\nmc{\swtwo}{s_\mathrm{w}^2}
\nmc{\cwtwo}{c_\mathrm{w}^2}
\nmc{\cwfour}{c_\mathrm{w}^4}
\nmc{\mw}{ {M_{\PW}}}
\nmc{\mwtwo}{ {M_{\PW}^2}}
\nmc{\dmwtwofin}{\delta {M_{\PW}^{2,\mathrm{fin}}}}
\nmc{\mz}{ {M_{\PZ}}}
\nmc{\mztwo}{ {M_{\PZ}^2}}
\nmc{\dmztwofin}{\delta {M_{\PZ}^{2,\mathrm{fin}}}}
\nmc{\mhl}{M_{\Hl}}
\nmc{\mhltwo}{M_{\Hl}^2}
\nmc{\dmhltwofin}{\delta {m_{\Hl}^{2,\mathrm{fin}}}}
\nmc{\mhh}{M_{\Hh}}
\nmc{\mhhtwo}{M_{\Hh}^2}
\nmc{\dmhhtwofin}{\delta {m_{\Hh}^{2,\mathrm{fin}}}}
\nmc{\mha}{M_{\Ha}}
\nmc{\mhatwo}{M_{\Ha}^2}
\nmc{\mhc}{M_{\Hpm}}
\nmc{\mhctwo}{M_{\Hpm}^2}
\nmc{\dmhctwofin}{\delta {M_{\Hpm}^{2,\mathrm{fin}}}}
\nmc{\dzhhhlms}{\delta Z_{\Hh\Hl}^\msbar}
\nmc{\dzhlhhms}{\delta Z_{\Hl\Hh}^\msbar}
\nmc{\dzgha}{\delta Z_{\GZ\Ha}}
\nmc{\dzhag}{\delta Z_{\Ha\GZ}}
\nmc{\dzghams}{\delta Z_{\GZ\Ha}^\msbar}
\nmc{\dzhagms}{\delta Z_{\Ha\GZ}^\msbar}
\nmc{\dzghcms}{\delta Z_{\PGpm\Hpm}^\msbar}
\nmc{\dzhcgms}{\delta Z_{\Hpm\PGpm}^\msbar}
\nmc{\dthl}{\delta t_{\Hl}}
\nmc{\dthh}{\delta t_{\Hh}}
\nmc{\dthlfin}{\delta t_{\Hl}^\mathrm{fin}}
\nmc{\dthhfin}{\delta t_{\Hh}^\mathrm{fin}}
\nmc{\thlhh}{t_{\PHl \PHh}}
\nmc{\thag} {t_{\PHa \GZ}}
\nmc{\thcg} {t_{\PHpm \PGpm}}
\nmc{\dZefin}{\delta {Z_{\mathrm{e}}^\mathrm{fin}}}

\newcommand{\npr}{\texttt{npr}}

\newcommand{\power}{\texttt{power}}

\newcommand{\pow}{\texttt{pow}}

\newcommand{\uv}{{\scriptscriptstyle \rm UV}}
\newcommand{\ms}{{\scriptscriptstyle \rm MS}}
\newcommand{\ir}{{\scriptscriptstyle \rm IR}}
\newcommand{\irr}{{\scriptscriptstyle \rm IR2}}

\nmc{\ii}{\mathrm{i}}

\nmc{\al}{\alpha}
\nmc{\be}{\beta}
\nmc{\dbefin}{\delta\beta^\mathrm{fin}}
\nmc{\tb}{t_\be}
\nmc{\tbtwo}{t_\be^2}
\nmc{\tbmsbmdts}{\delta t_{\be,\tsone}^\msbar}
\nmc{\tbmsbmts}{\delta t_{\be,\tstwo}^\msbar}
\nmc{\cab}{c_{\al\be}}
\nmc{\sab}{s_{\al\be}}
\nmc{\msb}{M_\mathrm{sb}}
\nmc{\sa}{s_\al}
\nmc{\ca}{c_\al}
\nmc{\cae}{\cos\al}
\nmc{\sbs}{s_\be}
\nmc{\sbstwo}{s_\be^2}
\nmc{\cbs}{c_\be}
\nmc{\cbstwo}{c_\be^2}
\nmc{\sas}{s_\al}
\nmc{\sastwo}{s_\al^2}
\nmc{\cas}{c_\al}
\nmc{\castwo}{c_\al^2}
\nmc{\sbe}{\sin\be}
\nmc{\satwo}{\sin^2\al}
\nmc{\sbeptwo}{\sin^2\be}
\nmc{\cbe}{\cos\be}
\nmc{\Msb}{M_{\rm sb}}
\nmc{\Msbtwo}{M_{\rm sb}^2}
\nmc{\dMsbtwofin}{\delta M_{\rm sb}^{2,\mathrm{fin}}}
\nmc{\dMsbtwomsbar}{\delta M_{\rm sb}^{2,\msbar}}
\nmc{\dMsbtwomsbarFJ}{\delta M_{\rm sb, FJTS}^{2,\msbar}}
\nmc{\dmtwomsbar}{\delta m_{12}^{2,\msbar}}
\nmc{\dMsbtwommsbar}{\delta M_{\rm sb}^{2,\dmtwomsbar}}
\nmc{\msbar}{{\overline{\mathrm{MS}}}}
\newcommand{\ts}{{\it FJ~Tadpole Scheme}}
\nmc{\tsone}{\mathrm{MDTS}}
\nmc{\tstwo}{\mathrm{MTS}}
\nmc{\tsthree}{\mathrm{FJTS}}

\definecolor{lightblue}{rgb}{0.122, 0.435, 0.698}
\newtcbox{\mdlbox}{nobeforeafter,colframe=lightblue,colback=lightblue!3!white,boxrule=0.5pt,arc=4pt,
  boxsep=0pt,left=6pt,right=6pt,top=2pt,bottom=3pt,tcbox raise base}
\newcommand{\SM}{SM}
\newcommand{\THDM}{2HDM}
\newcommand{\HS}{HSESM}

\newcommand{\MDLSUP}[1]{\begin{center}{\it Model support:}\; #1 \end{center}}

\newcommand{\MODELTHDM}{\MDLSUP{\mdlbox{\THDM}}}
\newcommand{\MODELHS}{\MDLSUP{\mdlbox{\HS}}}
\newcommand{\MODELTHDMHS}{\MDLSUP{\mdlbox{\THDM} \mdlbox{\HS}}}

\newcommand\dangersign[1][2ex]{%
  \renewcommand\stacktype{L}%
  \scaleto{\stackon[1.3pt]{\color{red}$\triangle$}{\tiny !}}{#1}%
}

\newcommand{\DeprecatedMethodsNoEffect}{\dangersign[3ex] These subroutines are
deprecated and calling any of them has no effect.}

\newcommand{\MethodNoSupportYet}{\dangersign[3ex] This subroutine is not supported yet and calling it has no effect.}
\newcommand{\MethodsNoSupportYet}{\dangersign[3ex] These subroutines are not supported yet and calling them has no effect.}

\newlength{\parwidth}
\newenvironment{insertion}
{\setlength{\parwidth}{\textwidth}%
\addtolength{\parwidth}{-\leftmargin}%
\addtolength{\parwidth}{-\labelsep}%
\addtolength{\parwidth}{-3em}%
\\[.3ex]\hspace*{3ex}\begin{minipage}[t]{\parwidth}}
{\end{minipage}\\[.2ex]}

\def\bei{\begin{itemize}}
\def\eei{\end{itemize}}
\def\bq{\begin{equation}}
\def\eq{\end{equation}}
\def\bqa{\begin{eqnarray}}
\def\eqa{\end{eqnarray}}

\def\refse#1{\mbox{Section~\ref{#1}}}
\def\refses#1{\mbox{Sections~\ref{#1}}}

\def\citere#1{\mbox{Ref.~\cite{#1}}}
\def\citeres#1{\mbox{Refs.~\cite{#1}}}
\def\refta#1{\mbox{Table~\ref{#1}}}

\def\reffi#1{\mbox{Fig.~\ref{#1}}}

\begin{document}

\begin{frontmatter}

\title{\RecolaTwo\\
REcursive Computation of One-Loop Amplitudes 2
\tnoteref{mytitlenote}
{\sc\normalsize Version 2.0}\\[-.2cm]
}
\tnotetext[mytitlenote]{The program is available from  
  \href{http://recola.hepforge.org/}{\mbox{http://recola.hepforge.org}}.}

\author[a]{Ansgar~Denner}
\ead{ansgar.denner@physik.uni-wuerzburg.de}
\author[a]{Jean-Nicolas~Lang}
\ead{jlang@physik.uni-wuerzburg.de}
\author[b]{Sandro~Uccirati}
\ead{uccirati@to.infn.it}

\address[a]{Universit\"at W\"urzburg, 
Institut f\"ur Theoretische Physik und Astrophysik, \\
D-97074 W\"urzburg, Germany}
\address[b]{Universit\`a di Torino e INFN, 10125 Torino, Italy\\[-.5cm]}

\begin{abstract}\sloppy
  We present the Fortran95 program \recolatwo\ for the perturbative
  computation of next-to-leading-order transition amplitudes in the
  Standard Model of particle physics and extended Higgs sectors. New
  theories are implemented via model files in the 't Hooft--Feynman
  gauge in the conventional formulation of quantum field theory and in
  the Background-Field method. The present version includes model
    files for the Two-Higgs-Doublet Model and the Higgs-Singlet Extension
    of the Standard Model. We support standard renormalization
  schemes for the Standard Model as well as many commonly used
  renormalization schemes in extended Higgs sectors. Within these
  models the computation of next-to-leading-order polarized amplitudes
  and squared amplitudes, optionally summed over spin and colour, is
  fully automated for any process.  \recolatwo{} allows the
  computation of colour- and spin-correlated leading-order squared
  amplitudes that are needed in the dipole subtraction formalism.
  \recolatwo\ is publicly available for download at
  http://recola.hepforge.org.
\end{abstract}

\begin{keyword}
NLO computations; 
one-loop amplitudes; Beyond Standard Model; higher orders; theories beyond the Standard Model
\end{keyword}

\end{frontmatter}

\clearpage

\section{Introduction}
\label{introduction}

In the era after the Higgs-boson discovery the focus in elementary
particle physics is on the precise validation of the Standard Model
(SM) and the search for possible extensions thereof a.k.a.~Beyond
Standard Model (BSM) theories. Nowadays we are able to perform
precision predictions in the SM for a vast number of observables as a
result of the automation of one-loop QCD
\cite{Hahn:2000kx,Agrawal:2012cv,Berger:2008sj,vanHameren:2009dr,Cullen:2011ac,
  Badger:2010nx,Badger:2012pg,Alwall:2014hca,Cascioli:2011va} and
electroweak (EW) \cite{Actis:2012qn, Kallweit:2014xda, Frixione:2015zaa,
  Chiesa:2015mya, Hahn:2000kx, Actis:2016mpe}
corrections. In the future, amplitude providers need to be able to
calculate one-loop QCD and EW corrections in general weakly
interacting theories. However, the automation of one-loop EW
corrections for BSM theories is more involved as several intermediate
steps are required.  The first step consists in the definition of new
models in terms of a Lagrangian and a subsequent determination of the
Feynman rules.  This can be done by means of \Feynrules{}
\cite{Christensen:2008py,Christensen:2009jx} and Sarah
\cite{Staub:2013tta}.  Then, the renormalization needs to be addressed
which raises many questions for BSM theories as parameters cannot be
necessarily linked to physical observables.  Thus, many different
renormalization schemes need to be investigated
which in turn requires a systematic and flexible approach. Steps
towards the automation of the renormalization for BSM theories have
been undertaken in the {\sc FeynRules}/{\sc FeynArts} approach in
\citere{Degrande:2014vpa}.  Finally, the renormalized model file needs
to be interfaced to a generic one-loop matrix-element generator. In
\citere{Denner:2017vms} we proposed a complementary strategy to
\citere{Degrande:2014vpa} combining the second and third step which
requires a small set of external tools (\FORM{}
\cite{Vermaseren:2000nd,Ruijl:2017dtg} and \Reptil{}
\cite{Denner:2017vms}).  Our approach makes use of tree-level {\sc
  Universal FeynRules Output} (UFO) model files \cite{Degrande:2011ua}
and results in renormalized one-loop model files for \recolatwo, a
generalized version of \recola, allowing anyone to compute any process
in the underlying theory at the one-loop level, with the only
restriction being available memory and CPU power. Much effort has been
spent on the validation of our system,
and for this purpose  the alternative formulation of
quantum field theory (QFT) in the Background-Field method (BFM) has
been implemented.
In this report we describe the \recolatwo{} library and the
\recolatwo{} model files for the computation of tree and one-loop
amplitudes in the SM, the Two-Higgs-Doublet Model (\THDM) and the
Higgs-Singlet Extension of the SM (\HS).

This article is organized as follows: In \refse{what-is-new-in-recola} we
summarize the new features of \recolatwo{} compared to the prior version
\recola~\cite{Actis:2016mpe}. In \refse{installation} the installation instructions for the
\recolatwo{} library and the model files are given. In \refse{calling recola} we
describe the usage of \recolatwo{} and comment on all new subroutines
related to models with 
extended Higgs sectors that can be called by the user.  Finally, we conclude in
\refse{conclusions} and list validation efforts in \ref{appendix_checks}.

\section{New features in \recolatwo\label{what-is-new-in-recola}}

\recolatwo{} is an upgraded version of the \FortranNinety{} code
\recola{}~\cite{Actis:2016mpe} for the computation of tree-level and one-loop
scattering amplitudes for general QFT, based on recursion
relations~\cite{Actis:2012qn}.  At tree level the algorithm computes amplitudes
using Dyson--Schwinger
equations~{\cite{Dyson:1949ha,Schwinger:1951ex,Schwinger:1951hq}}.
At one-loop order the recursion relies on the decomposition of one-loop
amplitudes in terms of tensor integrals, computed by means of the \collier{}
library \cite{Denner:2016kdg}, and tensor coefficients computed by \recolatwo,
within the framework of dimensional regularization.  The extension to BSM
concerns, in particular,  the computation of tensor coefficients as they are
process- and theory-dependent.  The \recolatwo{} library can generate arbitrary
processes in BSM theories and allows for the computation of tensor coefficients
involving new structures compared to the usual formulation of the \SM.

In \citere{Denner:2017vms} we have presented our algorithm for a fully
automated renormalization and computation of one-loop amplitudes. The
intermediate results of this approach are \recolatwo-specific
renormalized model files which are derived from nothing but tree-level
\UFO{} format \cite{Degrande:2011ua} by means of the tool \Reptil.
For the \THDM{} and \HS{} we specify in \refse{higgssector potentials}
and \refse{yukawa sector} the modified Higgs potentials and Yukawa
sector, respectively. The tree-level \UFO{} model files have been
derived using \Feynrules{}
\cite{Christensen:2008py,Christensen:2009jx}.  With extended Higgs
sectors the parameter space grows and requires renormalization of
additional parameters. In \refse{renormalization higgssectors} we list
all implemented renormalization schemes available in the model files.
In \refse{dimensional regularization} we give details on the
implementation of dimensional regularization and specify conventions
used in \msbar{} renormalization schemes.  In \refse{background field
  method} we comment on the formulation of model files in the BFM and,
finally, in \refse{conventions} we list our conventions for the new
fields used to define processes.

\subsection{Extended Higgs sectors and their renormalization}
\subsubsection{Scalar potentials}
\label{higgssector potentials}
Currently we support the \THDM{} and \HS{} as presented in
\citere{Denner:2017vms} and summarized in the following.  The models
with extended Higgs sector are distinguished from the \SM{} by
modified scalar potentials and Yukawa couplings.  We derived the
models for CP-conserving $Z_2$-symmetric renormalizable potentials.
Under these constraints the most general scalar potential in the
\THDM{} reads \cite{Gunion:2002}
\begin{align}
  V_{\mathrm{THDM}}={}&m_1^2\Phione^{\dagger}\Phione+m_2^2\Phitwo^{\dagger}\Phitwo
    -m_{12}^2\left(\Phione^{\dagger}\Phitwo+\Phitwo^{\dagger}\Phione\right)\notag\\
    &+\frac{\lambda_1}{2}\left(\Phione^{\dagger}\Phione\right)^2
    +\frac{\lambda_2}{2}\left(\Phitwo^{\dagger}\Phitwo\right)^2
    +\lambda_3\left(\Phione^{\dagger}\Phione\right)\left(\Phitwo^{\dagger}\Phitwo\right)\notag\\
    &+\lambda_4\left(\Phione^{\dagger}\Phitwo\right)\left(\Phitwo^{\dagger}\Phione\right)
    +\frac{\lambda_5}{2}\left[\left(\Phione^{\dagger}\Phitwo\right)^2
    +\left(\Phitwo^{\dagger}\Phione\right)^2\right],
\label{eq:thdmpot}
\end{align}
with $\Phi_1, \Phi_2$ being Higgs doublets. The five couplings
$\lambda_1\ldots\lambda_5$ and the two mass parameters $m^2_1$ and $m^2_2$ are
chosen to be real. Further, we allow for soft-breaking of the $Z_2$ symmetry which is
parametrized by the real parameter $m_{12}^2$.

Before spontaneous symmetry breaking (SSB) the parameters are
expressed in the symmetric, defining basis and by the EW gauge
couplings \g{} and \gy{} in the gauge eigenbasis.  \recolatwo{}
operates in the physical basis expressed by masses, mixing angles and
electromagnetic coupling. For the \THDM{} this results in the
following identification of physical parameters:
\begin{center}
\begin{tabular}{|l|l|l|}
  \hline
   basis & $V_{\mathrm{2HDM}}$ & $\mathcal{L}_{\mathrm{Gauge}}$ \\
   \hline
  \hline
  before SSB & $m_1, m_2, m_{12}, \lambda_1, \lambda_2, \lambda_3, \lambda_4, \lambda_5$ &
  $g, g'$ \\
  after SSB  & ${\mhl}, \mhh, \mha, \mhc, \cab,\tb, M_\mathrm{sb}, \mw$
  & $e, \mz$\\
   \hline
\end{tabular}
\end{center}
The masses are uniquely defined as the eigenvalues of the canonically normalized
mass matrices.
Here, $\mhl$ and $\mhh$ denote the light and heavy neutral Higgs-boson masses,
respectively, $\mha$ denotes the pseudo-scalar Higgs-boson mass and $\mhc$ the
charged Higgs-boson mass.  Besides being positive definite the neutral ones are
constraint to $\mhl <
\mhh$.
The angles $\alpha$ and $\beta$ are introduced to identify mass eigenstates and
Goldstone-boson degrees of freedom. We follow the conventions in
\citere{Gunion:2002} where the dependence on the angles is parametrized as
\begin{align}
  \alpha,\; \beta \quad\to \quad\cab:=\cos(\al-\be),\; \tb:= \tan \beta,
\end{align}
which is a natural choice for studying aligned scenarios. Here, $\tb$ is defined
as $\tb = \mathrm{vev}_2/\mathrm{vev}_1$, \ie the ratio of the vevs associated
to $\Phi_2$ and $\Phi_1$
\cite{Denner:2017vms,Denner:2016etu}. Below we give a selection of Feynman rules
which can be used to fix the conventions.
The angle $\alpha$ is defined in the window $[-\pi/2, \pi/2]$, whereas the
angle $\beta$ is defined in the window $[0, \pi/2]$. This implies $\ca: = \cae =
\sqrt{1-\satwo}$, $\cbs:= \cbe = \sqrt{1-\sbeptwo}$ and $\tb>0$.
Finally, we define the soft-breaking scale $\Msb$ as
\begin{align}
  \Msbtwo=\frac{m_{12}^2}{\cbs \sbs }.
\end{align}
We support two alternative sets of input parameters for mixing angles, namely:
\begin{center}
  \begin{tabular}{|l|l|}
     \hline
     parameter choice &  domain \\
     \hline
     \hline
     $ \begin{array}{c} \cab \\ \tb \end{array} $ &
     $ \begin{array}{c} [-1, 1] \\ (0, \infty) \end{array} $ \\
     \hline
     $ \begin{array}{c} \sa \\ \cbs \end{array} $ &
     $ \begin{array}{c} [-1, 1] \\ (0, 1) \end{array} $ \\
     \hline
  \end{tabular}
\end{center}

For the \HS{} the most general CP-conserving $Z_2$-symmetric renormalizable
scalar potential reads
\begin{align}
  V_{\mathrm{HSESM}} = m_1^2 \Phi^\dagger \Phi + m_2^2 S^2
  +\frac{\lambda_1}{2} \left(\Phi^\dagger \Phi\right)^2
  +\frac{\lambda_2}{2} S^4
  +\lambda_3 \Phi^\dagger \Phi S^2,
  \label{eq:hspot}
\end{align}
with $\Phi$ being a Higgs doublet and $S$ being a singlet field, and all
parameters are real. We choose the following set of physical parameters:
\begin{center}
\begin{tabular}{|l|l|l|}
  \hline
  basis & $V_{\mathrm{HSESM}}$ & $\mathcal{L}_{\mathrm{Gauge}}$ \\
   \hline
  \hline
  before SSB  &$m_1, m_2, \lambda_1, \lambda_2, \lambda_3$ &
  $g, g'$ \\
  after SSB  &${\mhl, \mhh}, \sas,\tb, \mw$ & $e, \mz$\\
  \hline
\end{tabular}
\end{center}
The angle $\alpha$ is defined in the same way as in the \THDM{} and $\tb$ is
defined as $\tb = \mathrm{vev}_s/\mathrm{vev}$, \ie the ratio of the vevs
associated to $S$ and $\Phi$ \cite{Denner:2017vms}.  Again, \recolatwo{}
operates in the physical basis expressed by masses, mixing angles and
electromagnetic coupling.

For comparison of phase conventions we list key couplings of type VVS ($g^{\mu
\nu}$ omitted) and SSS in \refta{ta:higgscouplings}.
\begin{table}
\begin{tabular}{|c|c|c|}
  \hline
  model & vertex & coupling \\
  \hline
  \hline \rule{0ex}{2.5ex}
  \THDM  & $\PZ \PZ \Hl$ & $-\ii \sab \frac{e\mz }{\cw \sw}$\\
         & $\PZ \PZ \Hh$ & $+\ii \cab \frac{e\mz }{\cw \sw}$\\
         & $\Ha \Ha \Hl$ & \footnotesize$+\ii \frac{e}{2 \mw \sw}
         \left( \left( 2\left( \mhatwo - \Msbtwo\right) +\mhltwo\right) \sab
         - \left(\mhltwo-\Msbtwo\right) \cab \frac{1-\tbtwo}{\tb} \right)$\\
         & $\Ha \Ha \Hh$ & \footnotesize$-\ii \frac{e}{2 \mw \sw}
         \left( \left( 2\left( \mhatwo - \Msbtwo\right) +\mhhtwo\right) \cab
         + \left(\mhhtwo-\Msbtwo\right) \sab \frac{1-\tbtwo}{\tb} \right)$\\
  \hline \rule{0ex}{2.5ex}
  \HS  & $\PZ \PZ \Hl$ & $-\ii \sas \frac{e\mz }{\cw \sw}$\\
       & $\PZ \PZ \Hh$ & $+\ii \cas \frac{e\mz }{\cw \sw}$\\
       & $\PHl \PHl \Hh$ & $-\ii \cas \sas \frac{ e}{2 \mw \sw \tb}
       \left( \mhhtwo + 2 \mhltwo\right) \left( \cas + \sas \tb\right) $\\
       & $\PHl \PHh \Hh$ & $-\ii \cas \sas \frac{ e}{2 \mw \sw \tb}
       \left( \mhltwo + 2 \mhhtwo\right) \left( \sas - \cas \tb\right) $\\
  \hline
\end{tabular}
\caption{Some key couplings in the \THDM{} and \HS.}
\label{ta:higgscouplings}
\end{table}
The SM limit is reached for $\sab=-1$ and $\sas=-1$ in the \THDM{} and 
\HS, respectively.
\subsubsection{Yukawa sector}
\label{yukawa sector}
While for the \HS{} the fermionic sector is the same as in the SM, the \THDM{}
allows for a richer structure with both doublet fields $\Phi_1$ and $\Phi_2$
coupling to fermions. Imposing a diagonal CKM matrix and the $Z_2$ symmetry,
which is essential for suppressing flavour-changing neutral currents at
tree level, leads to the natural flavour-conserving models.
The full Yukawa Lagrangian with all possible types reads
\begin{align}
  \mathcal{L}_\mathrm{Y}=
  &-\Gamma_\mathrm{d}  \overline{Q}_\mathrm{L}
  \left(h_{1,\mathrm{d}}\Phi_1 + h_{2,\mathrm{d}}\Phi_2\right)
  d_\mathrm{R}\notag\\
  &-\Gamma_\mathrm{u} \overline{Q}_\mathrm{L}
  \left(h_{1,\mathrm{u}}\tilde {\Phi}_1  + h_{2,\mathrm{u}}\tilde
  {\Phi}_2\right) u_\mathrm{R}\notag\\
  &-\Gamma_\mathrm{l} \overline{L}_\mathrm{L}
  \left(h_{1,\mathrm{l}}\Phi_1 + h_{2,\mathrm{l}}\Phi_2\right)l_\mathrm{R} 
  +\mathrm{h.c.},
\end{align}
with $\tilde{\Phi}_i$ being the charge conjugation of $\Phi_i$, and
$\Gamma_\mathrm{d}$, $\Gamma_\mathrm{u}$, $\Gamma_\mathrm{l}$ generically
denote the up-type quark, down-type quark and lepton Yukawa couplings.
The $\overline{Q}_\mathrm{L}$, $\overline{L}_\mathrm{L}$ and $d_\mathrm{R}$,
$u_\mathrm{R}$, $l_\mathrm{R}$ denote the SM fermion doublet and singlet fields,
respectively.
The parameters $h_{i, F}$ trigger the desired Yukawa type as follows:
\begin{center}
\begin{tabular}{|c|cccccc|}
  \hline 
  type & $h_{1,\mathrm{d}}$ & $h_{2,\mathrm{d}}$ & $h_{1,\mathrm{u}}$ &
  $h_{2,\mathrm{u}}$ & $h_{1,\mathrm{l}}$ & $h_{2,\mathrm{l}}$\\
  \hline 
  \hline 
  I  &  0 & 1 & 0 & 1 & 0 & 1 \\
  II &  1 & 0 & 0 & 1 & 1 & 0 \\
  X  &  0 & 1 & 0 & 1 & 1 & 0 \\
  Y  &  1 & 0 & 0 & 1 & 0 & 1 \\
  \hline
\end{tabular}
\end{center}
In this convention, the Yukawa couplings are generically given by
\begin{center}
  \begin{tabular}{|c|c|}
    \hline
    vertex & coupling\\
    \hline
    \hline \rule{0ex}{2.5ex}
    $\PHl F F$ & $-\ii \frac{m_{F} e}{2 \mw \sw \tb}
    \left(h_{2,{F}} \left(\cab - \sab \tb\right) -
          h_{1,{F}} \tb \left(\sab + \cab \tb\right)\right)$\\
    \hline \rule{0ex}{2.5ex}
    $\PHh F F$ & $-\ii \frac{m_{F} e}{2 \mw \sw \tb}
    \left(h_{2,{F}} \left(\sab + \cab \tb\right) +
          h_{1,{F}} \tb \left(\cab - \sab \tb\right)\right)$\\
    \hline
  \end{tabular}
\end{center}
with $h_{i,F}$ representing either $h_{i, \mathrm{d}}$,
$h_{i, \mathrm{u}}$ or $h_{i,\mathrm{l}}$ depending on the fermion type ${F}$.

\sloppypar
Note that the user can select the Yukawa type directly via
\texttt{set\_Z2\_thdm\_yukawa\_type\_rcl} (\refse{set yukawa type
thdm}), without having to worry about the values of $h_{i,F}$.

\subsubsection{Renormalization schemes}
\label{renormalization higgssectors}

The renormalization of the SM gauge couplings is performed as explained
in \citere{Denner:2017vms}. We support the renormalization of $\alpha$ in
the Thomson limit ($\alpha_0$) or on the Z-pole ($\alpha_\mathrm{Z}$).
Furthermore, $\alpha$ can be renormalized in the $G_{\rm F}$ scheme,
neglecting the muon mass in vertex and box contributions.

In the extended Higgs sectors the renormalization of mixing angles and
new couplings needs to be addressed.  We support various
renormalization schemes inspired by
\citeres{Denner:1990yz,Espinosa:2001xu,Espinosa:2002cd,
  Freitas:2002um,Sperling:2013eva,Bojarski:2015kra,Krause:2016oke,
  Denner:2016etu,Altenkamp:2017ldc}.  For all model files, except for
the \SM{} ones,\footnote{For the \SM{} model files we renormalize the
  tadpoles as done in \citere{Denner:1991kt}. On request we can
  provide the model files in different tadpole schemes.} the tadpoles
are by default renormalized in the \ts{} \cite{Fleischer:1980ub,
  Krause:2016oke, Denner:2016etu} which, roughly speaking, affects all
parameters depending on the vev. In particular, all counterterms to
physical parameters are gauge-parameter independent.  Nevertheless we
support other tadpole-counterterm schemes via appropriate shifts with
respect to counterterms defined in the \ts.  Note that the distinction
of different tadpole-counterterm schemes is only relevant for the
\msbar{} renormalization of $\alpha$, $\beta$ and $\Msb$, but not for
on-shell schemes or the \msbar{} renormalization of parameters of the
defining basis, \ie $\lambda_i, \mu_i$, because then the tadpoles
necessarily drop out.\footnote{There are subtleties for on-shell
  schemes defined in a particular gauge. See Appendix~C in
  \citere{Denner:2017vms} for a discussion and simple solution of this
  issue.} In summary, we distinguish between the following
tadpole-counterterm schemes present in the literature:
\begin{itemize}
  \item [\textbf{FJTS}:] In the {\bf F}leischer-{\bf J}egerlehner {\bf T}adpole counterterm
    {\bf S}cheme tadpole counterterms are introduced via field redefinitions.
    See Appendix~A in~\citere{Fleischer:1980ub}.
  \item [\textbf{MDTS}:] In the {\bf M}ass-{\bf D}iagonal {\bf T}adpole counterterm {\bf S}cheme
    no tadpole counterterms emerge for two-point functions of physical fields.
    See the Feynman rules in \citere{Denner:1991kt}.
  \item [\textbf{MTS}:] In the {\bf M}inimal {\bf T}adpole counterterm
    {\bf S}cheme 
    the translation to the physical basis is done in such a way that the 
    tadpole counterterms only appear in the quadratic terms of the Higgs
    potential.
    See Eq.~(6) in \citere{Actis:2006}.
\end{itemize}

The masses and fields of new (scalar) fields are renormalized in the on-shell
scheme in the same way as in the \SM. The only remaining parameters requiring
renormalization are mixing angles and the soft-breaking scale $\Msb$
in the \THDM.
Since $\alpha$, $\beta$  and $\Msb$ are considered as independent parameters (or
are related to direct derivatives thereof, \eg $\cab,\tb,\sas,\ldots$) we
formulate all renormalization schemes directly for the corresponding
counterterms $\delta \alpha$, $\delta \beta$ ($\delta \tb$) and $\delta
\Msbtwo$.
\begin{itemize}
\item \msbar{} schemes~\cite{Freitas:2002um,Sperling:2013eva,Denner:2016etu,Altenkamp:2017ldc}:\\
  We support standard \msbar{} schemes. In each of these schemes
  terms proportional to $\Delta_\uv$ are subtracted as explained
  in~\refse{dimensional regularization}. The UV finiteness of matrix
  elements can be tested by varying the scale $\mu_\uv$. The actual
  scale dependence of the scheme can be probed by varying the scale
  $\mu_\ms$.  The counterterms can be derived from suited vertices of
  the theory, or from the pole part of the off-diagonal field
  renormalization constants.
    \begin{description}
      \item[$\delta \alpha^\msbar$:]
        In our conventions for the \THDM\ and \HS{} 
        (see also \citere{Denner:2017vms}) we get for $\al$ in the \tsthree:
        \begin{align}
          \delta \alpha^\msbar_{\mathrm{FJTS}} := 
          \delta \alpha^\msbar & = \frac{\dzhhhlms-\dzhlhhms}{4},
          \label{da msbar}
        \end{align}
        which can be translated to other tadpole schemes as follows
        \begin{align}
          \delta \alpha^\msbar_{\tsone} &:= \delta
          \alpha^\msbar  + \frac{t_{\Hh \Hl}^\mathrm{fin}}{\mhhtwo-\mhltwo},
          \label{da msbar MDTS}
          \\
          \delta \alpha^\msbar_{\mathrm{MTS}} &:= \delta
          \alpha^\msbar  +
          \frac{t_{\Hh \Hl}^\mathrm{fin}-t_{\Hh \Hl,\mathrm{MTS}}^\mathrm{fin}}{\mhhtwo-\mhltwo},
          \label{da msbar MTS}
        \end{align}
        with $t_{\Hh \Hl}$ and $t_{\Hh \Hl,\mathrm{MTS}}$ being the
        tadpole counterterms to the neutral mixing energy in the FJTS and MTS
        tadpole counterterm schemes, respectively.

      \item[$\delta \lambda_3^\msbar$,
            $\delta \lambda_{345}^\msbar$:]
        Instead of $\alpha$ being renormalized
        $\msbar$, the user can choose between $\lambda_3$ and $\lambda_{345}$.
        In the \THDM{} this is equivalent to define the counterterm of $\alpha$
        in the following ways:\footnote{See also Eqs.~(4.38) and (4.39) in
        \citere{Altenkamp:2017ldc}.}
{\allowdisplaybreaks 
       \begin{align}
          &\delta \alpha^{\delta \lambda_{345}^\msbar} = \delta \alpha^\msbar
          - \frac{\cas \sas}{\castwo-\sastwo}2\dZefin
          +\frac{\cbstwo-\sbstwo}{\castwo-\sastwo}
          \frac{\cas \sas }{ \cbs \sbs}
          \dbefin
          \notag\\
          &-\frac{\cas \sas}{\castwo-\sastwo}
          \Bigl[
            \frac{\cwtwo-\swtwo}{\swtwo} \frac{\dmwtwofin}{\mwtwo}
            - \frac{\cwtwo}{\swtwo} \frac{\dmztwofin}{\mztwo}
            +\frac{\dmhhtwofin-\dmhltwofin}{\mhhtwo-\mhltwo}
          \Bigr]\notag\\
        &-\frac{\cbs \sbs}{\castwo-\sastwo}
          \Biggl[
            \frac{1}{\mhhtwo-\mhltwo}
          \Bigl[
              \dMsbtwofin\notag\\
            &+\Msbtwo
            \left( 
            2 \dZefin + 
            \frac{\cwtwo-\swtwo}{\swtwo} \frac{\dmwtwofin}{\mwtwo}
            - \frac{\cwtwo}{\swtwo} \frac{\dmztwofin}{\mztwo}
            \right)
          \Bigl]
          \Biggl],
        \label{da l345 msbar}
        \\
          &\delta \alpha^{\delta \lambda_3^\msbar} = \delta \alpha^{\delta \lambda_{345}^\msbar}\notag\\
        &-\frac{\cbs \sbs}{\castwo-\sastwo}
            \frac{2}{\mhhtwo-\mhltwo}
          \Biggl[
            \dmhctwofin -\dMsbtwofin\notag\\
            &+\left(
            \mhctwo-\Msbtwo
            \right)
            \left(
            \dZefin +
            \frac{\cwtwo-\swtwo}{\swtwo} \frac{\dmwtwofin}{\mwtwo}
            - \frac{\cwtwo}{\swtwo} \frac{\dmztwofin}{\mztwo}
            \right)
          \Biggl].
        \label{da l3 msbar}
        \end{align}}%
        In the \HS{} only the $\delta \lambda_3^\msbar$ scheme exists which can
        be defined via $\delta \alpha$ as follows:
        \begin{align}
          &\delta \alpha^{\delta \lambda_3^\msbar} = \delta \alpha^\msbar
          -\frac{\cas \sas}{\castwo-\sastwo} 2 \dZefin \notag \\
          &-\frac{\cas \sas}{\castwo-\sastwo}
          \Biggl[
            \frac{\cwtwo-\swtwo}{\swtwo} \frac{\dmwtwofin}{\mwtwo}
            - \frac{\cwtwo}{\swtwo} \frac{\dmztwofin}{\mztwo}
            +\frac{\dmhhtwofin-\dmhltwofin}{\mhhtwo-\mhltwo}
          \Biggr].
          \label{da l3 msbar HS}
        \end{align}

      \item[$\delta \beta^\msbar$:]
        In the \THDM\ we get for $\beta$ in the \tsthree:
        \begin{align}
          \delta \beta^\msbar_{\mathrm{FJTS}} := 
          \delta \beta^\msbar  & = \frac{\dzhagms-\dzghams}{4} =
          \frac{\dzhcgms-\dzghcms}{4},
          \label{db msbar}
        \end{align}
        while the results in the other tadpole counterterm schemes read
        \begin{align}
          \delta \beta^{\msbar}_{\tsone} &= 
          \delta \beta^{\msbar}_{\mathrm{MTS}} = 
          \delta \beta^{\msbar}
          + \left(\cab \frac{\dthlfin}{\mhltwo}
          + \sab \frac{\dthhfin}{\mhhtwo}\right) \frac{e}{2 \mw \sw}.
          \label{db TS msbar}
        \end{align}
        The \msbar{} scheme \eqref{db TS msbar} 
        corresponds to the popular $\overline {\mathrm{DR}}$ (\msbar)
        scheme in the Minimal Supersymmetric SM (see \eg
        \citere{Freitas:2002um}).  Note that the finite parts of the
        tadpoles are, in general, gauge dependent, and we fix them in
        the 't Hooft--Feynman gauge.  The \msbar/\tsone~scheme is used by
        \underline{default} for $\delta \beta$.
        
      \item[$\delta \tb^\msbar$:]
        In the \HS{} we renormalize $\tb$ in the \msbar{} scheme using the
        vertex $V_{\PHh \PHl \PHl}$ and require the pole part (P.P.) to vanish:
        \begin{align}
          \left. V_{\PHh \PHl \PHl}\right|_\mathrm{P.P.} \overset{!}{=} 0
          \quad\Rightarrow\quad \delta t^\msbar_{\be}=:\delta t^\msbar_{\be,\mathrm{FJTS}}.
          \label{dtb msbar HS}
        \end{align}
        For $\tb$ we offer the possibility to switch to the 
        \tsone{} scheme which is related to the \tsthree{} via
        \begin{align}
          \label{dtb TS msbar HS}
          \tbmsbmdts &= \delta \tb^{\msbar} \\
          &\quad {}+ \left(\left(\sas + \cas \tb\right) \frac{\dthlfin}{\mhltwo}
          -       \left(\cas - \sas \tb\right)
          \frac{\dthhfin}{\mhhtwo}\right)\frac{e}{2 \mw \sw}. \notag
        \end{align}
        We do not support the MTS for $\tb$ in the \HS.  The finite
        parts of the tadpoles are, in general, gauge dependent, and we
        fix them in the 't Hooft--Feynman gauge. The
          \msbar/\tsone~scheme is used by \underline{default} for
        $\delta \tb$ in the \HS.

      \item[$\dMsbtwomsbar$, $\dmtwomsbar$:]
        We determine $\dMsbtwomsbar$ from the vertex $V_{\PHh \PHpm \PHpm}$ 
        by requiring that the pole part (P.P.) vanishes:
        \begin{align}
          \left. V_{\PHh \PHpm \PHpm}\right|_\mathrm{P.P.} \overset{!}{=} 0
          \quad\Rightarrow\quad \dMsbtwomsbar=:\dMsbtwomsbarFJ.
          \label{dMsb msbar}
        \end{align}
        This scheme is
        the $\underline{\mathrm{default}}$.
        Alternatively, $m_{12}^2$ can be renormalized \msbar{} which is equivalent
        to define 
        \begin{align}
          \dMsbtwommsbar = \dMsbtwomsbar + 
          \Msbtwo \frac{\sbstwo - \cbstwo}{\cbs \sbs} \dbefin.
          \label{dm12 msbar}
        \end{align}
    \end{description}
  \item $p^*$ schemes
    \cite{Espinosa:2001xu,Espinosa:2002cd,Bojarski:2015kra,Krause:2016oke,Denner:2017vms}:\\
    The $p^*$ scheme is derived via the diagonalization of the
    effective mass matrix requiring vanishing scale dependence
    \cite{Espinosa:2001xu,Espinosa:2002cd}. In the \THDM, this
      scheme can be used also for the renormalization of the mixing
    angle $\beta$ yielding the following solutions:
    \begin{description}
      \item [$\delta \alpha^{p^*}$:] 
    \begin{align}
      \delta \alpha^{p^*} &= \frac{\Sigma^{\mathrm{1PI},\mathrm{BFM}}_{\Hh
      \Hl}\left(\frac{\mhhtwo+\mhltwo}{2}\right)+\thlhh}{\mhhtwo-\mhltwo},
      \label{da ps}
    \end{align}
    \item [$\delta \beta^{p^*_1}$:] 
    \begin{align}
      \delta \beta^{p^*_1} = -\frac{\Sigma^{\mathrm{1PI},\mathrm{BFM}}_{\Ha
      \GZ}\left(\frac{\mhatwo}{2}\right)+\thag}{\mhatwo},
      \label{db ps1}
    \end{align}
    \item [$\delta \beta^{p^*_2}$:] 
    \begin{align}
      \delta \beta^{p^*_2} = -\frac{\Sigma^{\mathrm{1PI},\mathrm{BFM}}_{\Hpm
      \PGpm}\left(\frac{\mhctwo}{2}\right)+\thcg}{\mhctwo}.
      \label{db ps2}
    \end{align}
    \end{description}
    Here, $\Sigma_{\Hh \Hl}$, $\Sigma_{\Ha \GZ}$ and $\Sigma_{\Hpm \PGpm}$
    denote the neutral, pseudo-scalar, and charged scalar mixing-energy,
    respectively, and $\thlhh$, $\thag$, $\thcg$ are the corresponding tadpole
    counterterms.
    The schemes are formulated via the BFM in the 't
    Hooft--Feynman gauge, \ie for the quantum gauge parameter
    $\xi_Q=1$ \cite{Denner:1994xt}.
    The scheme $\delta \alpha^{p^*}$ is valid in the \THDM{} and \HS{},
    whereas the schemes $\delta \beta^{p^*_1}$ and $\delta \beta^{p^*_2}$ can
    only be used in the \THDM.
  \item on-shell schemes \cite{Denner:1990yz,Krause:2016oke,Denner:2017vms}:\\
    In these schemes, the counterterms for the mixing angles are
    defined via on-shell mixing field-renormalization constants in the
    $\xi_Q=1$ gauge in the BFM.
    We express the counterterms in terms of mixing energies as follows
    \begin{description}
      \item [$\delta \alpha^\mathrm{OS}$:]
       \begin{align}
         \delta \alpha^\mathrm{OS} &= 
         \frac{\Sigma^{\mathrm{1PI},\mathrm{BFM}}_{\Hh \Hl}\left(\mhhtwo\right)+
       \Sigma^{\mathrm{1PI},\mathrm{BFM}}_{\Hl \Hh}\left(\mhltwo\right)  + 2
       \thlhh}{2\left(\mhhtwo-\mhltwo\right)},
        \label{da os}
      \end{align}
      \item [$\delta \beta^{\mathrm{OS}_1}$:]
        \begin{align}
          \delta \beta^{\mathrm{OS}_1}&=
         -\frac{\Sigma^{\mathrm{1PI},\mathrm{BFM}}_{\Ha\GZ}\left(0\right)+
         \Sigma^{\mathrm{1PI},\mathrm{BFM}}_{\Ha \GZ }\left(\mhatwo\right) + 2
         \thag}{2 \mhatwo},
        \label{db os1}
        \end{align}
      \item [$\delta \beta^{\mathrm{OS}_2}$:]
        \begin{align}
         \delta \beta^{\mathrm{OS}_2} &=
         -\frac{\Sigma^{\mathrm{1PI},\mathrm{BFM}}_{\Hpm\PGpm}\left(0\right)+
         \Sigma^{\mathrm{1PI},\mathrm{BFM}}_{\Hpm\PGpm}\left(\mhctwo\right) + 2
         \thcg}{2 \mhctwo}.
        \label{db os2}
        \end{align}
    \end{description}
    The scheme $\delta \alpha^\mathrm{OS}$ is valid in the \THDM{} and \HS,
    whereas the schemes $\delta \beta^{\mathrm{OS}_1}$ and $\delta
    \beta^{\mathrm{OS}_2}$ can only be used in the \THDM.
\end{itemize}

\subsection{\msbar{} renormalization and scale dependence}
\label{dimensional regularization}
\recolatwo{} distinguishes between poles in $\epsilon$ of infrared (IR) and
ultraviolet (UV) origin by introducing
the parameters $\mu_\uv$, $\epsilon_\uv$ and $\mu_\ir$,
$\epsilon_\ir$ in all tensor integrals and counterterms, together with
\bqa
\Delta_\uv &=& 
\frac{(4\pi)^{\epsilon_\uv}\,\Gamma(1+\epsilon_\uv)}{\epsilon_\uv},
\nonumber\\
\Delta_\ir &=& 
\frac{(4\pi)^{\epsilon_\ir}\,\Gamma(1+\epsilon_\ir)}{\epsilon_\ir},
\qquad
\Delta_\irr \;=\; 
\frac{(4\pi)^{\epsilon_\ir}\,\Gamma(1+\epsilon_\ir)}{\epsilon_\ir^2}.
\eqa
Following the conventions of
\collier~\cite{Denner:2014gla,Denner:2016kdg}
and \citere{Denner:2010tr}, the parameters $\Delta_\uv$, $\Delta_\ir$
and $\Delta_\irr$ that contain the poles in $\epsilon$ absorb a
normalization factor of the form $1+{\cal O}(\epsilon)$.  In terms of
these parameters, the one-loop amplitude ${\cal A}_1$ takes the general form
\bq
{\cal A}_1 = 
  \Delta_\uv\,{\cal A}_1^\uv
+ \Delta_\irr\,{\cal A}_1^\irr
+ \Delta_\ir\,{\cal A}_1^\ir(\mu_\ir)
+ {\cal A}_1^{\rm fin}(\mu_\uv,\mu_\ir).
\eq
The term ${\cal A}_1^\uv$ vanishes after (UV) renormalization. The a
priori unphysical scale $\mu_\uv$ should either cancel between the
tensor integrals and the counterterms, or it should receive a physical
interpretation as for example when it is identified with the
renormalization scale $\mu_\ms$ in the $\overline{\rm MS}$
scheme\footnote{In the original \recola{} version as well in the
  literature the scale $\mu_\ms$ is called $Q$.} for the strong
coupling constant $g_s$.  In \recolatwo, the \msbar{} renormalization
scale $\mu_\ms$ is introduced as a new independent scale by a modified
\msbar{} renormalization where instead of $\Delta_\uv$ the term
\begin{align}
  \Delta_\uv + \ln \frac{\mu_\uv^2}{\mu_\ms^2}
  \label{eq:defmsbsub}
\end{align}
is subtracted in \msbar{} schemes. As a consequence, the renormalized one-loop
amplitude becomes
\bq
{\cal A}_1 = 
  \Delta_\irr\,{\cal A}_1^\irr
+ \Delta_\ir\,{\cal A}_1^\ir(\mu_\ir)
+ {\cal A}_1^{\rm fin}(\mu_\ms,\mu_\ir)
\eq
which is independent of $\mu_\uv$ but can depend on $\mu_\ms$ when parameters
are renormalized \msbar. Note that in the extended Higgs sectors besides the
strong coupling constant $g_s$ additional sources of scale dependence emerge.
The numerical evaluation of the renormalized amplitude involves parts which
depend on $\Delta_\uv$ and $\mu_\uv$ at intermediate steps.
The independence of ${\cal A}_1$ on $\Delta_\uv$ and $\mu_\uv$ can be
verified numerically by varying these parameters.

\subsubsection{Soft and collinear singularities \label{collinear}}
In \recolatwo{} collinear singularities are
  regularized as in \recola{} according to the input fermion
    masses which are forwarded to \collier, while soft singularities
  are always regularized dimensionally.\footnote{In \recola{} the
    regularization of soft singularities is steered by the parameter
    \texttt{reg\_soft}.  Associated input routines can still be called
    in \recolatwo, but are deprecated and have no effect.}  In the
  case of massless fermions dimensional regularization is used, and
  the scale dependence of regularized integrals is parametrized by the
  IR scale $\mu_\ir$.  If, on the other hand, the fermion has been
  assigned a regulator mass, its collinear singularities are
  regularized by this mass parameter.  In this case, the parameter
  $\mu_\ir$ can be interpreted as a photon-mass regulator.

\subsection{Background-Field Method}
\label{background field method}

\begin{figure}
\begin{center}
\begin{picture}(120, 170)(-60,-160)
\Text(3,9)[cc]{\scriptsize $  32\;\;{\scriptstyle{\rm g}_{Q}}$}
\Gluon(0,0)(0, -40){1.5}{9}
\Text(-45, -20)[rc]{\scriptsize   1}\Text(-35, -20)[lc]{\scriptsize ${\scriptstyle{\rm Z}}$}
\Photon(-20, -20)(0, -40){1.5}{ 7}
\Text(-45, -40)[rc]{\scriptsize   2}\Text(-35, -40)[lc]{\scriptsize ${\rm u}$}
\ArrowLine(-20, -40)(0, -40)
\Text(-45, -60)[rc]{\scriptsize   4}\Text(-35, -60)[lc]{\scriptsize $\bar{\rm u}$}
\ArrowLine(0, -40)(-20, -60)
\Gluon(0, -40)(20, -65){1.5}{8}
\Text(  14, -51)[lb]{\scriptsize ${\scriptstyle{\rm g}_{Q}}$}
\GCirc(0, -40){3}{.5}
\Text(-45, -80)[rc]{\scriptsize   8}\Text(-35, -80)[lc]{\scriptsize ${\rm g}$}
\Gluon(-20, -80)(0, -90){1.5}{ 5}
\Text(-45,-100)[rc]{\scriptsize  16}\Text(-35,-100)[lc]{\scriptsize ${\rm g}$}
\Gluon(-20,-100)(0, -90){1.5}{ 5}
\Gluon(0, -90)(20, -65){1.5}{ 8}
\Text(  14, -85)[lb]{\scriptsize ${\rm g}$}
\GCirc(0, -90){3}{.5}
\Gluon(20, -65)(40, -65){1.5}{5}
\Text(46, -65)[lc]{\scriptsize ${\scriptstyle{\rm g}_{Q}}$}
\GBoxc(20, -65)(6,6){1}
\Text(-50,-114)[l]{\scriptsize Incoming Colour Structures:}
\Text(-47,-126)[lb]{\scriptsize $  39$}
\Text(-30,-125)[l]{\scriptsize $\delta_{j2}^{i32}\delta_{j32}^i\delta_j^{i4}$}
\Text(-47,-138)[lb]{\scriptsize $  24$}
\Text(-30,-137)[l]{\scriptsize $\delta_{j16}^{i8}\delta_{j8}^i\delta_j^{i16}$}
\Text(-50,-149)[l]{\scriptsize Outgoing Colour Structure:}
\Text(-30,-160)[l]{\scriptsize $\delta_{j2}^{i16}\delta_{j8}^{i4}\delta_{j16}^{i8}$}
\SetWidth{1.5}
\Line(-3,-3)(3,3)\Line(-3,3)(3,-3)
\Line(37, -68)(43, -62)\Line(37, -62)(43, -68)
\end{picture}%
\end{center}%
\caption{Visualization of a branch with a quantum field.}%
\label{fi:BFMcurrent}
\end{figure}
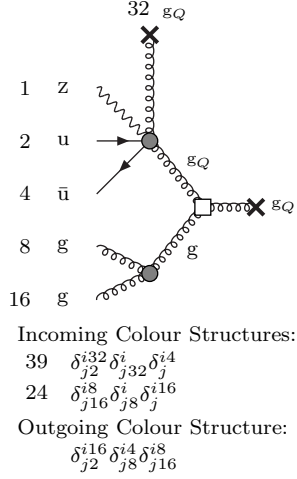

\recolatwo{} supports the Background-Field Method (BFM)
as a complementary method to the usual formulation of
  QFT. The implementation is realised for each individual model as a
separate model file.  Whether a model file is formulated in the BFM is
printed in the \texttt{Gauge} entry in the initialisation of the
model. For example, the \SM{} BFM initialisation reads:

{\footnotesize
\begin{verbatim}
xxxxxxxxxxxxxxxxxxxxxxxxxxxxxxxxxxxxxxxxxxxxxxxxxxxxxxxxxxxxxxxxxxxxxxxxxxx
                        _    _   _   _        _    _ 
                       | )  |_  |   | |  |   |_|   _|
                       | \  |_  |_  |_|  |_  | |  |_

               REcursive Computation of One Loop Amplitudes

                               Version 2.0.0

                     by A.Denner, J.-N.Lang, S.Uccirati

xxxxxxxxxxxxxxxxxxxxxxxxxxxxxxxxxxxxxxxxxxxxxxxxxxxxxxxxxxxxxxxxxxxxxxxxxxx

---------------------------------------------------------------------------
 Active model:     SM
 Gauge:            't Hooft-Feynman BFM
 Model generation: Tue Aug 29 18:58:55 2017
---------------------------------------------------------------------------
\end{verbatim}
}
The BFM and the usual formulation of QFT are treated on equal footing, and, from
the user perspective, models formulated in the BFM can be used as any other
model since the same conventions for background fields in the BFM and the usual
fields (\refse{conventions}) were chosen.
The BFM comes with additional fields, so-called quantum fields, which cannot be
selected as external states of matrix elements because they only appear inside
loops. When drawing the currents to one-loop scattering processes the quantum
fields can be seen at intermediate steps. For instance, the propagation of a
gluon quantum field ${\rm g}_Q$ along the loop line (the line distinguished by
the cross at the beginning and end) is visualized in \reffi{fi:BFMcurrent}.
Note that for all model files formulated in the BFM the fermion fields take the
role of both, quantum and background field and no explicit distinction is made
\cite{Denner:2017vms}.

All renormalization schemes available in the usual formulation are
also available in the BFM allowing for powerful cross checks.  There
are, however, (higher-order) differences for mass counterterms due
  to the expansion performed within the complex-mass scheme (see
  \citere{Denner:2017vms}).  For a precise numerical comparison of
  both methods it is recommended to set the widths of all particles to
  zero where these differences disappear.  In practical calculations
  the BFM formulation is $10{-}15\%$ faster than the conventional
  formulation which is due to the need of less current structures.  On
  the other hand, the generation of processes takes $\sim50\%$ longer
  in the BFM.

\subsection{Conventions}
\label{conventions}

\recolatwo\ uses the same conventions for particle identifiers (of
type {\tt character}) as \recola{} \cite{Actis:2016mpe} with the
exception of the Goldstone bosons:\\[1.5ex]
\quad\begin{tabular}{ll} Scalars in the HSESM: & \texttt{ 'Hl', 'Hh' }
  \\
  Scalars in the THDM: & \texttt{ 'Hl', 'Hh', 'Ha', 'H+', 'H-' }
  \\
  Goldstone bosons: & \texttt{ 'G0', 'G+', 'G-' } \; .
\end{tabular}
\\[1.5ex]
Other \SM{} fields as well as conventions for polarizations and
the normalization of the cross section are given in \citere{Actis:2016mpe}.

\section{Installation}
\label{installation}

\recolatwo\ uses the \collier\ library and \recolatwo-specific model
files. In order to facilitate building \recolatwo{} we provide
the following two options:
\bei
\item \recolatwo-\collier\ package (\refse{recola collier model-file
  package}):\\
The configuration and compilation of \recolatwo, \collier, and a model
file is performed at once, using a \CMake{} script which resolves all
dependencies automatically. 

\item \recolatwo-stand-alone package (\refse{recola package}):\\ A stand-alone
  version which requires the user to resolve the dependence to \collier{} and
  model files by hand.
\eei

The \recolatwo{} library and model files are available from the web site 
\url{http://recola.hepforge.org}.
The compilation requires the \textsc{CMake} build system.

\subsection{The \recolatwo-\collier\ package}
\label{recola collier model-file package}

The package \texttt{recola2-collier-X.Y.Z} contains
the version \texttt{X.Y.Z} of the \recolatwo\ library together with all model
files.
After downloading 
the file \texttt{recola2-collier-X.Y.Z.tar.gz}, extract the tarball in the 
current working directory with the shell command:
\begin{verbatim}
   tar -zxvf recola2-collier-X.Y.Z.tar.gz
\end{verbatim}
This operation creates the directory \texttt{recola2-collier-X.Y.Z} containing
the following files and folders:
\bei
\item\texttt{CMakeLists.txt}, \texttt{build}:\\
  \CMake{} configuration for the compilation of the \recolatwo{},  \collier{}
  and model-file libraries in the proper order. The build directory, where
  \CMake{} puts all necessary files for the creation of the library;
\item\texttt{recola2-X.Y.Z}:\\
main directory of the \recolatwo\ package
\texttt{recola2-X.Y.Z} (see \refse{recola package} for 
details);
\item\texttt{model-files-X.Y.Z}:\\
directory containing the \recolatwo\ model files:
\begin{itemize}
  \item \texttt{SM\_X.Y.Z}, \texttt{SM\_BFM\_X.Y.Z}
  \item \texttt{HS\_X.Y.Z}, \texttt{HS\_BFM\_X.Y.Z}
  \item \texttt{THDM\_X.Y.Z}, \texttt{THDM\_BFM\_X.Y.Z}
\end{itemize}
Each model carries the same version as
the \recolatwo{} library;
\item \texttt{project\_cmake}:\\
  a test configuration which demonstrates how to link \recolatwo{} to an external
  project using \CMake;
  this is discussed in \refse{test program cmake}.
\eei
By running
\begin{verbatim}
  cd recola2-collier-X.Y.Z/build
  cmake [options] .. -Dmodel=<model>
  make [options]
\end{verbatim}
the \recolatwo{} library is compiled and linked with the model \texttt{<model>} and the
\collier{} library.
Note that \texttt{..} is the relative path pointing from the build directory one level
higher to where the top-level CMakeLists.txt of the recola2-collier package is located.
A predefined version of the \collier{} library is
automatically downloaded\footnote{
  \collier{} is downloaded from
\href{http://collier.hepforge.org/}{\mbox{http://collier.hepforge.org}}
  If the \collier{} source is already present no download step is performed.}
and extracted next to the \recolatwo{} sources upon invoking the \texttt{make}
command.
The variable \texttt{<model>} can be one of the model ids \texttt{SM},
\texttt{SM\_BFM}, \texttt{HS}, \texttt{HS\_BFM}, \texttt{THDM},
\texttt{THDM\_BFM}, or it can be an absolute or relative path pointing to
the model-file sources.
Selecting a different model file requires to rerun the \CMake{} configuration
and compilation with a different value for \texttt{model}.
The \texttt{[options]} for the configuration (\texttt{cmake}) and compilation
(\texttt{make}) are explained in \refse{recola package} and summarized in
\refta{ta:settings summary}.  It is recommended to run the compilation
parallelised with the Makefile option \texttt{-j}.  Once the compilation of
\recolatwo{} is finished the demo files (see \refse{demo files}) can be run.
\begin{table}
\footnotesize
\begin{tabular}{|c|c|c|}
  \hline
  \CMake{} option      & Value  & Short description\\
  \hline
  \hline
  \texttt{collier\_path} & Path & Absolute or relative path to the  \\ 
  & & \collier{} library.\\
  \hline
  \texttt{modelfile\_path} & Path & Absolute or relative path to the  \\
  & & \recolatwo{} model file. Only available in\\
  & & \recolatwo{} \texttt{CMakeLists.txt}. \\
  \hline
  \texttt{static} & On/Off & Compile the library as a shared \\
  & & or static library.\\
  \hline
  \texttt{with\_python3} & On/Off & Choose \PythonThree{} over \PythonTwo{}\\
  & & to compile \texttt{pyrecola}. Only available in\\
  & & \recolatwo{} \texttt{CMakeLists.txt}. \\
  \hline
  \texttt{with\_smtests} & On/Off & Run tests against \pole{} and \openloops.\\
  & & Only available in \recolatwo{}\\
  & & \texttt{CMakeLists.txt}. \\
  \hline
  \texttt{CMAKE\_BUILD\_TYPE} & Debug/Release & Set the compiler flags. By
  default \\
  & &Release flags (optimized) are used.\\
  \hline
  \texttt{CMAKE\_Fortran\_COMPILER} & Path/Name & Set the \Fortran{} compiler
  either \\
  & & via executable name or the absolute\\
  & & path to executable.\\
  \hline
  \texttt{CMAKE\_INSTALL\_PREFIX} & Path & Set the installation prefix.\\
  \hline
  \hline
  Makefile option & Value  & Short description\\
  \hline
  \hline
  \texttt{-j} &  Integer & Number of threads for compilation. \\
  \hline
  \texttt{VERBOSE} &  True/False & Enable \textit{verbose} compilation. In this\\
   & & mode all compilation flags are\\
   & & visible to the user.\\
  \hline
\end{tabular}
\caption{Summary of the \CMake{} and Makefile options.}
\label{ta:settings summary}
\end{table}

\subsubsection{The \recolatwo{} demo files}
\label{demo files}
The \recolatwo{} demo files are located in \texttt{recola2-X.Y.Z}/demos.
In order to compile and run executables for the demo programs the command
\begin{verbatim}
   ./run <demofile>
\end{verbatim}
can be executed with \texttt{<demofile>} being either \texttt{demo0\_rcl},
\texttt{demo1\_rcl}, \texttt{demo2\_rcl}, \texttt{demo3\_rcl},
\texttt{demo4\_rcl}, or \texttt{demo5\_rcl}.
Alternatively, the user can execute
\begin{verbatim}
   make <demofile>
\end{verbatim}
in the build directory \texttt{recola2-X.Y.Z/build} to compile the respective
executable.

The demo programs exemplify the
usage of \recolatwo\ for various purposes:
\bei
\item \texttt{demo0\_rcl}:\\
Basic usage of \recolatwo.
\item \texttt{demo1\_rcl}:\\
  Usage of \recolatwo\ for more than one process simultaneously, with
  explicit modification of input parameters and with selection of
  specific helicities for the external particles and of certain powers
  of the strong coupling constant.  In addition, files with \LaTeX\ 
  source code for diagrams are generated.
\item \texttt{demo2\_rcl}:\\
Usage of \recolatwo\ for the selection of resonant contributions 
and pole approximation.
\item \texttt{demo3\_rcl}:\\
\sloppy
Usage of \recolatwo\ for the computation of colour- and/or 
spin-correlation.
\item \texttt{demo4\_rcl}:\\
\sloppy
Usage of \recolatwo\ for the computation of decay widths using the example of light
 and heavy neutral Higgs decays into Higgs and gauge bosons: 
 \bei
 \item $\Hh \to \Hl \Hl$,
 \item $\Hh/\Hl \to \PZ \PZ$, \quad $\Hh/\Hl \to \PZ \gamma$, \quad $\Hh/\Hl \to \gamma \gamma$,
 \item $\Hh/\Hl \to \Pg \Pg$.
 \eei
 Moreover, the use of different renormalization schemes in the extended Higgs
 sector is demonstrated.
\item \texttt{demo5\_rcl}:\\
  \sloppy Usage of \recolatwo\ for the selection of powers of
  coupling types in the new system at the example of a vector-boson-fusion
  partonic channel.  \eei Note that the demo files
  \texttt{demo0\_rcl}, \texttt{demo1\_rcl}, \texttt{demo2\_rcl},
  \texttt{demo3\_rcl} are identical to the ones in \recola, but can be
  used with any current \recolatwo{} model file.

For each \texttt{<demofile>} corresponding \Cpp (\texttt{cdemo}) and \Python{}
(\texttt{pydemo}) demo files are available. The \Cpp{} demo files are compiled in the same
way, substituting \texttt{<demofile>} by the precise \texttt{cdemo} file name.
The \Cpp{} interface is described in \refse{calling recola from cpp}.
The \Python{} demo file can be run directly without compilation. See 
\refse{calling recola from python} for more details on the \Python{} interface.

The {\tt demos} directory also contains the shell script \texttt{draw-tex} which
compiles all \LaTeX\ files of the form \texttt{process\_*.tex} present in the
folder and creates the corresponding \texttt{.pdf} files. 
It can be run by invoking 
\begin{verbatim}
   ./draw-tex
\end{verbatim}
in the {\tt demos} directory. 
\recolatwo{} provides the same functionally for drawing
currents as \recola{} and we refer to \citere{Actis:2016mpe} for details.

\subsubsection{A minimal executable with \CMake}
\label{test program cmake}
The purpose of this section is to demonstrate how to link \recolatwo{} to
a custom program using the \CMake{} build system.
The \recolatwo-\collier{} package comes with a \FortranNinety{} and \Cpp{} test
program in the directory \texttt{project\_cmake}.
We discuss the \FortranNinety{} version, \texttt{program.f90}, which merely contains the lines of code:
\begin{verbatim}
  program main
    use recola
  end program
\end{verbatim}
That is, the program includes \recolatwo{} and can be extended by the user at
will.
For the compilation the essential part is the \CMake{} configure file
\texttt{CMakeLists.txt}. A template \texttt{CMakeLists.txt} is provided:
\begin{verbatim}
  cmake_minimum_required(VERSION 2.8)
  project(example Fortran C CXX)

  add_executable(program program.f90)

  find_package(recola
               "X.Y.Z" EXACT REQUIRED
               HINTS "../install")

  target_link_libraries(program ${RECOLA_LIBRARY_PATH})
\end{verbatim}
The first two lines define a new project and declare \Fortran{}
as the main language.\footnote{Change the compiler language to \CC{} via 
\texttt{project(example C)}. Multiple languages at the same time are allowed, 
\eg \texttt{project(example Fortran C CXX)}. \CMake{} automatically decides which
compiler to take based on the suffix of files.}
The third statement adds a new executable named \texttt{program} which is
compiled from the source \texttt{program.f90}. Since the program uses
\recolatwo{} it needs to be linked against it. This is achieved in two steps. In
the first step, the \recolatwo{} package is searched for via the \CMake{}
command \texttt{find\_package}.
The version requirement \texttt{exact} can be relaxed by removing
it and omitting the minor/patch version.
On success, this command fills the \CMake{} variables
\bei
\item \texttt{RECOLA\_LIBRARY\_DIR}: \\
  directory containing the \recolatwo{} library;
\item \texttt{RECOLA\_INCLUDE\_DIR}: \\
  \sloppy
  directory containing the \recolatwo{} header files, \ie the
  \Fortran{} modules \texttt{*.mod}, the \CC{} header file \texttt{recola.h}, and
  the \Cpp header file \texttt{recola.hpp};
\item \texttt{RECOLA\_LIBRARY\_PATH}:\\
  \sloppy
  the absolute path to the \recolatwo{} library file
  \texttt{librecola.so} (or \texttt{librecola.a}, \texttt{librecola.dynlib}).
\eei
On failure, the configuration stops printing an error.
After successfully finding the \recolatwo{} package the
executable is linked with the \recolatwo{} library.

Note this script works independent of whether \recolatwo{} is compiled as shared or
static library, whether the underlying program is \Fortran{}, \CC{} or \Cpp{}
based, and independent of the underlying operating system or compilers.
For further tuning we refer to \url{https://cmake.org/Wiki/CMake}.

\subsection{The \recolatwo\ stand-alone package}
\label{recola package}
In this section we discuss the compilation of the stand-alone \recolatwo{}
library, the compilation of the model files, and all available compiler options.
We assume the user has acquired a local copy of \collier
\footnote{\collier\ can be downloaded from
\href{http://collier.hepforge.org/}{\mbox{http://collier.hepforge.org}}.}
and that he/she followed the \collier{} compilation instructions.

The archives
\bei
  \item \texttt{recola2-X.Y.Z.tar.gz} 
  \item \texttt{SM\_X.Y.Z.tar.gz}, \texttt{SM\_BFM\_X.Y.Z.tar.gz}
  \item \texttt{THDM\_X.Y.Z.tar.gz}, \texttt{THDM\_BFM\_X.Y.Z.tar.gz}
  \item \texttt{HS\_X.Y.Z.tar.gz}, \texttt{HS\_BFM\_X.Y.Z.tar.gz}
\eei
are available at \url{http://recola.hepforge.org} and represent
the \recolatwo{} library and model files in the version \texttt{X.Y.Z}.

\subsubsection{The \recolatwo\ model-file compilation\label{model compilation}}
In the current version of \recolatwo{} the model files are a dependency at
compile time,\footnote{Dynamic loading of model files at run-time is not
supported.} thus, the desired model needs to be compiled first.
To this end, extract one of the model-file tarballs with the shell command:
\begin{Verbatim}[commandchars=\\\{\}]
   tar -zxvf {\it modelfile}-X.Y.Z.tar.gz
\end{Verbatim}
This operation creates the directory \textit{modelfile}\texttt{-X.Y.Z}
containing the following files and folders:
\bei
\item \texttt{CMakeLists.txt}, \texttt{config}, \texttt{build}: \\
\CMake{} configure files required for the generation of the \recolatwo\ model
file Makefile. The build directory is where \CMake{} puts all necessary
files for the creation of the library;
\item \texttt{src}: \\
model-file source directory;
\item \texttt{include}: \\
all \Fortran{} module
files \texttt{*.mod} are placed inside this folder.
\eei 

The compilation of the model file proceeds by changing to the \texttt{build}
directory\footnote{Changing to the build directory is optional but 
recommended. \CMake{} populates the working directory with many 
configuration and object files. With a build directory those 
files can be cleaned easily by removing the contents of the build folder.} 
and executing there the shell command \texttt{"cmake [options] .."} (creating 
a Makefile in 
\textit{modelfile}\texttt{-X.Y.Z/build}),
followed by \texttt{make}:
\begin{Verbatim}[commandchars=\\\{\}]
   cd {\it modelfile}-X.Y.Z/build
   cmake [options] .. 
   make [options]
\end{Verbatim}
Note that \texttt{..} is the relative path pointing from the build directory one level
higher to where the top-level CMakeLists.txt of the model file is located.
By default, the configuration will search for the \collier{} library in
directories next to the model-file directory. It is possible to force the use of
a particular \collier{} version by providing a path.
The \CMake{} variable \texttt{collier\_path} can be used to pass the
(absolute/relative) path to the directory containing the \collier{} library:
\begin{Verbatim}
   -Dcollier_path=<PATH_TO_COLLIER>
\end{Verbatim}
By using this option it is assumed that the \collier{} module files are located
inside the path:
\begin{Verbatim}
  <PATH_TO_COLLIER>/modules
\end{Verbatim}

As a third alternative the environment variables \texttt{COLLIER\_INCLUDE\_DIR}
and \texttt{COLLIER\_LIBRARY\_DIR} can be set to the directories including the
\collier{} modules and \collier{} library, respectively. In this case the
complete sequence of calls reads:
\begin{Verbatim}[commandchars=\\\{\}]
   cd {\it modelfile}-X.Y.Z/build
   export COLLIER_LIBRARY_DIR=<PATH_TO_COLLIER_LIBRARY>
   export COLLIER_INCLUDE_DIR=<PATH_TO_COLLIER_MODULES>
   cmake [options] ..
   make [options]
\end{Verbatim}
We stress that the library paths should not point to the precise \collier{} library
file, but only to the directory containing the library.

If no other options are specified, \CMake{} automatically searches for installed
\Fortran{} compilers and chooses a suited one.
The user can force \CMake{} to use a
specific compiler by appending to the \texttt{cmake} command the 
option
\begin{verbatim}
   -DCMAKE_Fortran_COMPILER=<comp>
\end{verbatim}
where \texttt{<comp>} can be \texttt{gfortran, ifort, pgf95, ...} or the 
full path to a compiler.

By default, the installation sequence generates the model file as a shared
library \texttt{libmodelfile.so(/.dynlib)} in the directory
\textit{modelfile}-\texttt{X.Y.Z}, with the corresponding module files placed in
the \texttt{include} 
subdirectory. The option
\begin{verbatim}
   -Dstatic=ON
\end{verbatim}
causes \textsc{CMake} to create the static library instead of the shared one.

After the configuration the compilation can be run parallelised which is
recommended as the model files contain plenty of independent source files.
Parallelised compilation is performed via
\begin{verbatim}
   make -j <THREADS>
\end{verbatim}
where \texttt{<THREADS>} is the number of parallel compilation units. 
The \CMake{} Makefile allows the compilation command to be run in \textit{verbose} mode via:
\begin{verbatim}
  make VERBOSE=True
\end{verbatim}
If desired, the model file can be installed into the system via the command:
\begin{verbatim}
   make install
\end{verbatim}
A default install prefix is set automatically by \CMake,\footnote{The default
installation path is set to inside the root system and requires root rights.}
but can be altered by configuring the Makefile via
\begin{verbatim}
  -DCMAKE_INSTALL_PREFIX=<INSTALLATION_PATH>
\end{verbatim}
with \texttt{<INSTALLATION\_PATH>} being a custom installation path, \eg
\texttt{\$HOME}.

\subsubsection{The \recolatwo\ library compilation}
The configuration and compilation of \recolatwo{} proceeds in the very same way
as for model files, with the exception that, in addition, a specific model-file
path has to be set.

In the first step, extract the \recolatwo{} tarball with the shell command:
\begin{Verbatim}[commandchars=\\\{\}]
   tar -zxvf recola2-X.Y.Z.tar.gz
\end{Verbatim}
This operation creates the directory \texttt{recola2-X.Y.Z}
containing the following files and folders:
\bei
\item \texttt{CMakeLists.txt}, \texttt{config}, \texttt{build}: \\
  \CMake{} configure files required for the generation of the \recolatwo\
  Makefile. The build directory is where \textsc{CMake} puts all necessary
  files for the creation of the library;
\item \texttt{src}: \\
  \recolatwo\ source directory;
\item \texttt{demos}: \\
  directory with demo programs illustrating the use of \recolatwo, including
  shell scripts for their compilation and execution;
\item \texttt{include}: \\
  directory with \CC{} and \Cpp{} header files; all \Fortran{} module
  files \texttt{*.mod} are placed inside this folder;
\eei 

The standard sequence for the compilation reads
\begin{Verbatim}[commandchars=\\\{\}]
   cd recola2-X.Y.Z/build
   cmake [options] .. -Dmodelfile_path=<PATH_TO_MODELFILE> 
                      -Dcollier_path=<PATH_TO_COLLIER>
   make [options]
\end{Verbatim}
where \texttt{<PATH\_TO\_MODELFILE>} and \texttt{<PATH\_TO\_COLLIER>} is the
path to the directory containing the compiled model-file and \collier{} library,
respectively.
By using this option for \collier{} it is assumed that the \collier{} module files are located
inside the path:
\begin{Verbatim}
  <PATH_TO_COLLIER>/modules
\end{Verbatim}

Alternatively, the environment variable \texttt{MODELFILE\_PATH} can be set to
the directory including the model file \CMake{} configure
files\footnote{\CMake{} generates configure files for built or installed model
files. They are named \texttt{modelfileConfig.cmake} and
\texttt{modelfileConfigVersion.cmake} and are required for linking to the
\recolatwo{} library.}
or the model-file library. In this case
the sequence of calls reads:
\begin{Verbatim}[commandchars=\\\{\}]
   cd recola2-X.Y.Z/build
   export MODELFILE_PATH=<PATH_TO_MODELFILE_CONFIG>
   cmake [options] .. -Dcollier_path=<PATH_TO_COLLIER>
   make [options]
\end{Verbatim}
We support the same \texttt{[options]} for configuration and compilation of
\recolatwo{} as for the model files.  Note also that the paths to the \collier{}
library and module files can be set via environment variables as described in
\refse{model compilation}.

Starting with \recolatwo{} we support a \Python{} interface. By default, the
\CMake{} configuration will search for the \PythonTwo{} library, header and
executable and will, on success, build the library \texttt{pyrecola} (see
\refse{calling recola from python}). We support \PythonThree{} which can be
enabled in the configuration via 
\begin{verbatim}
  -Dwith_python3=On
\end{verbatim}
and, evidently, requires the \PythonThree{} library, header and executable to be
present in the system. We note that the \recolatwo{} build is not
affected if the \Python{} dependencies are not fulfilled.

Finally, \recolatwo{} is equipped with a few test routines allowing to check the
proper integration of \recolatwo{} into the system. In order to be able to
compile the tests the configuration needs to be run with:
\begin{verbatim}
  -Dwith_smtests=On
\end{verbatim}
Then, after building \recolatwo, the tests can be run via 
\begin{verbatim}
  make test
\end{verbatim}
or by invoking
\begin{verbatim}
  ctest
\end{verbatim}
in the directory \texttt{recola2-X.Y.Z/build}. Note that the tests should only
be run with a SM model file, otherwise the tests will fail.

\section{Usage of \recolatwo{} in extended Higgs sectors}
\label{calling recola}
In order to use \recolatwo\ in a {\sc Fortran} program its modules
have to be loaded by including the line
\begin{verbatim}
  use recola
\end{verbatim}
\sloppy
in the preamble of the respective code, and the library {\tt librecola.so}, {\tt
librecola.dynlib} or {\tt librecola.a} has to be supplied to the linker.  This
gives access to the public functions and subroutines of the \recolatwo\ library
described in the following subsections. The names of all these routines end with
the suffix ``{\tt \_rcl}''. This name convention is supposed to avoid conflicts
with routine names present in the master program and increases readability by
allowing for an easy identification of command lines referring to the
\recolatwo\ library.

Typically, an application of \recolatwo\ involves the following five steps:

\bei
\item{\bf Step 1: Setting input parameters (optional)}
  
The input needed for the computation of processes can be set by calling
dedicated subroutines as provided by \recolatwo. See \refse{input subroutines}
for the additional set of methods related to extended Higgs sectors. 
Since \recolatwo\ provides default values for all input parameters,
this first step is optional.

\item{\bf Step 2: Defining the processes}

Before \recolatwo\ can be employed to calculate matrix elements 
for one or more processes, each process must be declared and labelled with
a unique identifier. This is done by calling the 
\sloppy
subroutine \texttt{define\_process\_rcl} for every process, as described
in \citere{Actis:2016mpe}.  The functionality of the process definition has been
extended which is documented in \refse{definition} for subroutines that are
concerned.
\item{\bf Step 3: Generating the processes}
 
  In the next step the subroutine \texttt{generate\_processes\_rcl} is
  called which triggers the initialisation of the complete list of
  processes defined in step 2.  As a result, all relevant building
  blocks for the recursive computation of off-shell currents are
  generated (see \citere{Actis:2016mpe} for details).
\item{\bf Step 4: Computing the processes}
  
  The computation of the amplitude and of the squared amplitude is
  performed by means of the subroutine \texttt{compute\_process\_rcl},
  which uses the process-dependent information on the recursive
  procedure derived in step 3. The subroutine
  \texttt{compute\_process\_rcl} is called with the momenta of the
  external particles provided by the user.  In a Monte Carlo
  integration, the call of \texttt{compute\_process\_rcl} is repeated
  many times for different phase-space points.


Detailed information on the subroutines that can be employed in step 4 can be
found in \citere{Actis:2016mpe}. The functionality of the process computation
has been extended which is documented in \refse{computation} for subroutines
that are concerned.
\item{\bf Step 5: resetting \recolatwo}

Finally, by calling the subroutine \texttt{reset\_recola\_rcl}, the
process-depen\-dent information generated in steps 2--4 is deleted and
the corresponding memory is deallocated. The input variables keep
their values defined in step 1 before.
\eei
Note that these steps have to be followed in the order given above. 
In particular, after step 3 no new
process can be defined unless \recolatwo\ is reset (step 5). After
step 5 the user can restart with step 1 or step 2. 
More information on the allowed sequence of calls can be found in
\citere{Actis:2016mpe}.

Examples are found in the directory \texttt{demos} and are described in
\refse{demo files}. 

\subsection{Input subroutines for parameters of extended Higgs sectors
\label{input exthiggs}}
The following input subroutines are complementary subroutines to the original
ones in \recola{} and can be used to set input parameters and renormalization
schemes in extended Higgs sectors. 
Note that these subroutines can only be called if supported by the selected
model file, otherwise \recolatwo{} prints an error message and stops.
\subsubsection{\tt set\_pole\_mass\_hl\_hh\_rcl (ml,gl,mh,gh)} 
This subroutine sets the pole masses and widths (in GeV) of the
light ($\PHl$) and heavy ($\PHh$) Higgs bosons 
to \ml, \gl{} and \mh, \gh, respectively (\ml, \gl, \mh{} and \gh{} are of type 
{\tt real(dp)}). The pole masses must fulfil the condition \ml $<$ \mh.                     
Note that the degenerate mass scenario is not supported.
\MODELTHDMHS

\subsubsection{\tt set\_pole\_mass\_ha\_rcl (m,g)} 
This subroutine sets the pole mass and width (in GeV) of the pseudo scalar
Higgs boson ($\PHa$) to \m\ and \g, respectively (\m\ and \g\ are of type {\tt
real(dp)}).
\MODELTHDM

\subsubsection{\tt set\_pole\_mass\_hc\_rcl (m,g)} 
This subroutine sets the pole mass and width (in GeV) of the charged 
Higgs boson ($\PHpm$) to \m\ and \g, respectively (\m\ and \g\ are of type {\tt
real(dp)}).
\MODELTHDM

\subsubsection{\tt set\_Z2\_thdm\_yukawa\_type\_rcl (ytype)}
\label{set yukawa type thdm}
This subroutine sets the Yukawa type in the softly-broken Z2 symmetric
\THDM{} to \texttt{ytype}.
The variable \texttt{ytype} is of type {\tt integer} and accepts the following values:
\begin{center}
\begin{tabular}{|l|l|l|}
  \hline
  \texttt{ytype} & label & zero parameters\\
  \hline
  \hline
  \texttt{1} & Type-I   & h1u=h1d=h1l=0 \\
  \texttt{2} & Type-II  & h1u=h2d=h2l=0 \\
  \texttt{3} & Type-X   & h1u=h1d=h2l=0 \\
  \texttt{4} & Type-Y   & h1u=h2d=h1l=0\\
  \hline
\end{tabular}
\end{center}
See \refse{yukawa sector} for more information.
\MODELTHDM

\subsubsection{\tt set\_tb\_cab\_rcl (tb,cab)\label{set_tb_cab_rcl}}
This subroutine sets the value of $\tb$ to \texttt{tb}, and $\cab$  to
\texttt{cab} [\texttt{tb} and \texttt{cab} are of type {\tt real(dp)}].
The value of \texttt{tb}\ is strictly greater than zero and the value of
\texttt{cab} must fulfil $-1 \le \texttt{cab} \le 1$.
If these conditions are violated an error is raised.
The renormalization for $\cab$ and $\tb$ is fixed via the renormalization of
$\al$ and $\be$ which can be set using the subroutines 
{\texttt{use\_mixing\_alpha\_}{\it rs}\texttt{\_scheme\_rcl}}
(\refse{set al reno}) and {\texttt{use\_mixing\_beta\_}{\it
rs}\texttt{\_scheme\_rcl}} (\refse{set
be reno}),
respectively.  See \refse{renormalization higgssectors} for more details.
\MODELTHDM

\subsubsection{\tt use\_mixing\_alpha\_{\it rs}\_scheme\_rcl ({\it s})}
\label{set al reno}
These subroutines ({\it rs}=\texttt{msbar},\texttt{onshell}) set the
renormalization scheme for the mixing angle $\al$ or a derivative thereof
to \textit{s} (of type {\tt character}).
The following \msbar{} and on-shell schemes are supported:
\begin{center}
\begin{tabular}{|l|l|l|}
  \hline
  \textit{rs} & \textit{s} & renormalization-scheme description \\
  \hline
  \hline \rule{0ex}{2.5ex}
  \!\!\texttt{msbar}   & \texttt{'FJTS'}   & $\al$ renormalized \msbar, FJ tadpole scheme \eqref{da msbar} \\
  \texttt{msbar}   & \texttt{'MDTS'}   & $\al$ renormalized \msbar, MD tadpole scheme \eqref{da msbar MDTS} \\
  \texttt{msbar}   & \texttt{'MTS'}  & $\al$ renormalized \msbar, minimal tadpole scheme \eqref{da msbar MTS} \\
  \texttt{msbar}   & \texttt{'l3'}   & $\lambda_3$ renormalized \msbar{}
  \eqref{da l3 msbar},\eqref{da l3 msbar HS} \\
  \texttt{msbar}   & \texttt{'l345'} & $\lambda_{345}$ renormalized \msbar{} \eqref{da l345 msbar} \\
  \hline \rule{0ex}{2.5ex}
  \!\!\texttt{onshell} & \texttt{'ps'}   & $\delta \al$ defined in the $p^*$ scheme \eqref{da ps} \\   
  \texttt{onshell} & \texttt{'os'}   & $\delta \al$ defined in the on-shell scheme \eqref{da os} \\
  \hline
\end{tabular}
\end{center}
Note that only one of the schemes can be used at a time. The \msbar{} scheme 
\texttt{'l345'} can only be used with the \THDM.
For more information on the schemes consider \refse{renormalization
higgssectors}.
\MODELTHDMHS

\subsubsection{\tt use\_mixing\_beta\_{\it rs}\_scheme\_rcl ({\it s})}
\label{set be reno}
These subroutines ({\it rs}=\texttt{msbar},\texttt{onshell}) set the
renormalization scheme for the mixing angle $\be$
to \textit{s} (of type {\tt character}).
The following \msbar{} and on-shell schemes are supported:
\begin{center}
\begin{tabular}{|l|l|l|}
  \hline
  \textit{rs} & \textit{s} & renormalization-scheme description \\
  \hline
  \hline \rule{0ex}{2.5ex}
  \!\!\texttt{msbar}    & \texttt{'FJTS'} & $\be$ is renormalized \msbar, FJ tadpole scheme \eqref{db msbar} \\
  \texttt{msbar}    & \texttt{'MDTS'}   & $\be$ is renormalized \msbar, MD tadpole scheme \eqref{db TS msbar} \\
  \hline \rule{0ex}{2.5ex}
  \!\!\texttt{onshell}  & \texttt{'ps1'} & $\delta\be$ defined in $p^*$ scheme via mixing $\PHa \PG$ \eqref{db ps1} \\   
  \texttt{onshell}  & \texttt{'ps2'} & $\delta\be$ defined in $p^*$ scheme via mixing $\PHpm \PGpm$ \eqref{db ps2}\\
  \texttt{onshell}  & \texttt{'os1'} & $\delta\be$ defined in on-shell scheme via mixing $\PHa \PG$ \eqref{db os1}\\
  \texttt{onshell}  & \texttt{'os2'} & $\delta\be$ defined in on-shell scheme via mixing $\PHpm \PGpm$ \eqref{db os2}\\   
  \hline
\end{tabular}
\end{center}
Note that only one of the schemes can be used at a time.
For more information on the schemes consider \refse{renormalization
higgssectors}.
\MODELTHDM

\subsubsection{\tt set\_msb\_rcl (msb)}
\label{set msb}
This subroutine sets the value of the soft-$Z_2$-breaking scale $\msb$ to
\texttt{msb} (\texttt{msb} is of type {\tt real(dp)}).
The renormalization scheme for $\msb$ is set with \texttt{use\_msb\_{\it
rs}\_scheme\_rcl} (\refse{set msb reno}).
See \refse{higgssector potentials} for more details.
\MODELTHDM

\subsubsection{\tt use\_msb\_msbar\_scheme\_rcl({\it s})}
\label{set msb reno}
This subroutine sets the
renormalization scheme for soft-$Z_2$-breaking scale $\msb$ or the
soft-$Z_2$-breaking parameter $m_{12}$ to \textit{s} (of type {\tt character}).
The following schemes are supported:
\begin{center}
\begin{tabular}{|l|l|}
  \hline
  \textit{s} & renormalization-scheme description \\
  \hline
  \hline \rule{0ex}{2.5ex}
  \!\!\texttt{'MSB'} & $\msb$ is renormalized \msbar{} \eqref{dMsb msbar}  \\
  \texttt{'m12'} & $m_{12}$ is renormalized \msbar{} \eqref{dm12 msbar}\\
  \hline
\end{tabular}
\end{center}
Note that only one of the schemes can be used at a time.
The input value of $\msb$ is set with \texttt{set\_msb\_rcl} (\refse{set
msb}).
See \refse{renormalization higgssectors} for more details.
\MODELTHDM

\subsubsection{\tt set\_sa\_rcl (sa)}
\label{set sa}
This subroutine sets the value of $\sa$ to \texttt{sa} [\texttt{sa} is of type
{\tt real(dp)}]. The input value of \texttt{sa} must fulfil
$-1 \le \texttt{sa} \le 1$, otherwise an error is raised.
The renormalization scheme for $\sa$ is fixed via the renormalization of $\al$
which can be set with \texttt{use\_mixing\_alpha\_{\it rs}\_scheme\_rcl} (\refse{set
al reno}).
See \refse{higgssector potentials} for more details.
\MODELHS

\subsubsection{\tt set\_tb\_rcl (tb)}
\label{set tb}
This subroutine sets the value of $\tb$ to \texttt{tb} [\texttt{tb} is of type
{\tt real(dp)}]. The passed value of \texttt{tb} must fulfil
$\texttt{tb} > 0$, otherwise an error is raised.
The renormalization scheme for $\tb$ is set with \texttt{use\_tb\_{\it
rs}\_scheme\_rcl} (\refse{set tb reno}).
See \refse{higgssector potentials} for more details.
\MODELHS

\subsubsection{\tt use\_tb\_msbar\_scheme\_rcl ({\it s})}
\label{set tb reno}
This subroutine sets the renormalization scheme for $\tb$ to \textit{s} (of type
{\tt character}). The following schemes are supported:
\begin{center}
\begin{tabular}{|l|l|}
  \hline
  \textit{s} & renormalization-scheme description \\
  \hline
  \hline \rule{0ex}{2.5ex}
  \!\!\texttt{'FJTS'} & $\tb$ is renormalized \msbar, FJ tadpole scheme \eqref{dtb msbar HS} \\
  \texttt{'MDTS'}   & $\tb$ is renormalized \msbar, MD tadpole scheme \eqref{dtb TS msbar HS} \\
  \hline
\end{tabular}
\end{center}
Note that only one of the schemes can be used at a time.
The input value of $\tb$ is set with \texttt{\tt set\_tb\_rcl} (\ref{set
tb}).
See \refse{renormalization higgssectors} for more details.
\MODELHS

\subsection{Compatibility with \recola{} input subroutines}
\label{input subroutines}

In this section we list new subroutines, original ones with modified
behaviour, and no longer supported ones that are kept for backward
compatibility.  We stress that all subroutines existing in \recola{}
can be called in \recolatwo.

\subsubsection{\tt use\_gfermi\_scheme\_rcl (g,a)\label{sec::usegf}}
\dangersign[3ex]
This subroutine (see Section 4.2.18 in \citere{Actis:2016mpe}) has a different
behaviour when being used in combination with the optional argument
\g\ which represents the Fermi constant $G_{\rm F}$.
Internally, $\g$ is translated to $\al$ according to
\begin{align}
  \al = \frac{\sqrt{2}\g \mwtwo}{\pi} \left(1-\frac{\mwtwo}{\mztwo}\right),
\end{align}
with the currently active values for the (real) vector-boson masses $\mw,\mz$.
Therefore, the vector-boson masses should be set before 
calling this subroutine.

\subsubsection{\tt set\_parameter\_rcl(param,value)\label{set_parameter_rcl}}
This subroutine sets the value of \texttt{param} (of type \texttt{character}) to
\texttt{value} (of type \texttt{complex(dp)}).  Note that only independent
couplings, masses and widths can be set via this subroutine, without any
consistency checks being performed.  An imaginary part of \texttt{value} can
lead to undefined behaviour, and \texttt{value} should be real even though it is
of type \texttt{complex(dp)}. The allowed values for \texttt{param} depend on
the model file and can be looked up in the subroutine
\texttt{set\_parameter\_mdl} defined in \texttt{class\_particles.f90} of the
respective model file. For example, setting the light Higgs-boson mass \mhl{} to
$125 \GeV$ in the \THDM{} or \HS{} is achieved as follows:
\begin{verbatim}
  call set_parameter_rcl("MHL", complex(125d0, 0d0))
\end{verbatim}
For extended Higgs sectors the dedicated subroutines
in \refse{input exthiggs} should be used to set input parameters.
When calling \texttt{set\_parameter\_mdl} with arguments that are
incompatible with the selected model file, a warning is printed.
The program does not stop, and the call has no effect.

\subsubsection{\tt set\_renoscheme\_rcl(ctparam,renoscheme)}
This subroutine sets the renormalization scheme for the parameter with name
\texttt{param} (of type \texttt{character}) to \texttt{renoscheme} (of type
\texttt{character}). The allowed values for \texttt{ctparam} and
\texttt{renoscheme} depend on the model file and can be looked up in the
subroutine \texttt{set\_renoscheme\_mdl} defined in
\texttt{fill\_ctparameters.f90} of the respective model file.  For example,
setting the renormalization scheme of the mixing angle $\alpha$ to the $p^*$
scheme \eqref{da ps} in the \THDM{} or \HS{} is achieved as follows:
\begin{verbatim}
  call set_renoscheme_rcl("da_QED2", "ps_bfm")
\end{verbatim}
For extended Higgs sectors the dedicated subroutines in \refse{input exthiggs}
should be used to set renormalization schemes.
When calling \texttt{set\_renoscheme\_mdl} with arguments that are
incompatible with the selected model file,  a warning is printed.
The program does not stop, and the call has no effect.

\subsubsection{\tt use\_dim\_reg\_soft\_rcl,\\
                   use\_mass\_reg\_soft\_rcl (m),\\
                   set\_mass\_reg\_soft\_rcl (m)
}
\DeprecatedMethodsNoEffect\ The regularization of light fermions is determined
automatically.  See \refse{collinear} for details.

\subsubsection{\tt set\_complex\_mass\_scheme\_rcl,\\
  set\_on\_shell\_scheme\_rcl} \DeprecatedMethodsNoEffect\ Note that
the model files are all derived in the complex-mass scheme.

\subsubsection{\tt set\_dynamic\_settings\_rcl (n)}
\MethodNoSupportYet

\subsubsection{\tt set\_print\_level\_parameters\_rcl (n)}
This subroutine, which can be called before the process generation but
also during the process computation, sets internal variables
governing the output of input, derived and counterterms parameters:
\bei
\item
\texttt{n = 0}: No parameters are printed.
\item
\texttt{n = 1}: Input parameters are printed.
\item
\texttt{n = 2}: Input and derived parameters are printed.
\item
\texttt{n = 3}: Input, derived and counterterm parameters are printed.
\eei
By default, only input parameters are printed.

\subsubsection{\tt set\_print\_level\_RAM\_rcl (n)}

\MethodNoSupportYet

\subsubsection{\tt scale\_coupling3\_rcl (fac,pa1,pa2,pa3),\\
                   scale\_coupling4\_rcl (fac,pa1,pa2,pa3,pa4),\\
                   switchoff\_coupling3\_rcl (pa1,pa2,pa3),\\
                   switchoff\_coupling4\_rcl (pa1,pa2,pa3,pa4)}

\MethodsNoSupportYet

\subsubsection{\tt set\_collier\_output\_dir\_rcl (dir)}
This subroutine, which can be called before the process generation, sets the
\collier{} output directory to \texttt{dir} (\texttt{dir} is of type {\tt
character}), with \texttt{dir} being a relative or absolute path.
The default \collier{} output directory can be enforced by passing \texttt{dir} =
\texttt{'default'}.


\subsection{Updates on process definition}
\label{definition}
In this section we describe the new treatment of powers of types of
fundamental couplings in
\recolatwo.  In theories with a SM gauge group structure amplitudes
are proportional to $g_s^{n_s}{e}^{n-n_s}$, where $g_s$ and $e$
represent the strong and electroweak gauge couplings. For a given
process and power $n_s$ in $g_s$ the power $n$ is unambiguously
determined. For this special case of theories we support the original
\recola{} methods \texttt{select\_gs\_power\_*} where only the powers
in $g_s$ are selected,\footnote{See Sections 4.3.2--4.3.6 in
  \citere{Actis:2016mpe}.} but for more general theories the user has
to employ the new general selection methods given in \refses{sel pow
  born} and \ref{sel pow loop}, exclusively available in \recolatwo.

The treatment of different coupling types is kept general and uses the
information on coupling powers as defined by model files in the \UFO{}
format.  For all model files we choose the powers in the strong and EW
gauge couplings as follows:
\begin{center}
  \begin{tabular}{|c|c|c|}
    \hline
    coupling type & position & coupling constant \\
    \hline
    \hline
    \texttt{'QCD'} & 1 & $g_s$\\
    \texttt{'QED'} & 2 & $e$\\
    \hline
\end{tabular}
\end{center}
The position is relevant only when calling a subroutine which
requires the couplings powers as an array of integers, \eg the subroutines in
\refse{computation}. For the selection of powers a string identifier is used as
described in the following.

\subsubsection{\tt select\_power\_BornAmpl\_rcl (npr,cid,power), \newline
               \texttt{unselect\_power\_BornAmpl\_rcl (npr,cid,power)}}%
\label{sel pow born}
This pair of subroutines allows to select/unselect the contribution to the Born
amplitude proportional to general powers in coupling types $c_{id}^n$ of
the underlying theory. Here, $c_{id}$ is set to \texttt{cid} (of type {\tt
character}), $n$ is given by the {\tt integer} argument \power, and
\npr\ is the process identifier (of type {\tt integer}).
The variable \texttt{cid} accepts the following values:
\begin{insertion}
\begin{tabular}{@{}ll@{}}
\texttt{'QCD'}: & power in $g_s$, \\
\texttt{'QED'}: & power in $e$.
\end{tabular}
\end{insertion}

All other contributions to the Born amplitude keep their status
(selected or unselected), according to previous calls of selection
subroutines.  The selection of the contributions to the loop amplitude
remains unaffected as well.  New values for \texttt{cid} will be
introduced in the future for theories with additional types of
couplings (as appear, for instance, in theories with new gauge
couplings or in effective field theories).  For \SM-like theories
with the only two fundamental coupling types \texttt{'QCD'} and \texttt{'QED'}
the methods \texttt{select\_gs\_power\_BornAmpl\_rcl} and
\texttt{unselect\_gs\_power\_BornAmpl\_rcl} (see
\citere{Actis:2016mpe}) can be used instead.

\subsubsection{\tt select\_power\_LoopAmpl\_rcl (npr,cid,power), \newline
               \texttt{unselect\_power\_LoopAmpl\_rcl (npr,cid,power)}}%
\label{sel pow loop}
This pair of subroutines allows to select/unselect the contribution to the loop
amplitude proportional to general powers in coupling types $c_{id}^n$ of
the underlying theory.
Here, $c_{id}$ is set to \texttt{cid} (of type {\tt
character}), $n$ is given by the {\tt integer} argument \power, and
\npr\ is the process identifier (of type {\tt integer}).
The variable \texttt{cid} accepts the following values:
\begin{insertion}
\begin{tabular}{@{}ll@{}}
\texttt{'QCD'}: & power in $g_s$, \\
\texttt{'QED'}: & power in $e$.
\end{tabular}
\end{insertion}
All other contributions to the loop amplitude keep their status (selected or 
unselected), according to previous calls of selection subroutines. 
The selection of contributions to the Born amplitude remains 
unaffected as well.
New values for \texttt{cid} will be introduced in the future for theories
with additional (gauge) couplings or in effective field theory.
For \SM-like theories with the only two fundamental coupling types \texttt{'QCD'} and
\texttt{'QED'} the methods 
\texttt{select\_gs\_power\_LoopAmpl\_rcl} and
\texttt{unselect\_gs\_power\_LoopAmpl\_rcl} (see \citere{Actis:2016mpe}) can be used instead.

\subsubsection{\tt select\_all\_gs\_powers\_BornAmpl\_rcl (npr), \newline
               unselect\_all\_gs\_powers\_BornAmpl\_rcl (npr), \newline
               select\_all\_powers\_BornAmpl\_rcl (npr), \newline
               unselect\_all\_powers\_BornAmpl\_rcl (npr)
               }
These subroutines allow to select/unselect all contributions to 
the Born amplitude (with any power of $g_s$ or general order) for the process with 
identifier \npr\ (of type {\tt integer}).
The selection of contributions to the loop amplitude remains 
unaffected. The methods with $g_s$ in their name have the same
effect and are only kept for backward compatibility.
In fact, all methods (un-)select \underline{all} contributions.

\subsubsection{\tt select\_all\_gs\_powers\_LoopAmpl\_rcl (npr), \newline
               unselect\_all\_gs\_powers\_LoopAmpl\_rcl (npr), \newline
               select\_all\_powers\_LoopAmpl\_rcl (npr), \newline
               unselect\_all\_powers\_LoopAmpl\_rcl (npr)}
This pair of subroutines allows to select/unselect all contributions to 
the loop amplitude (with any power of $g_s$ or general order) for the process with 
identifier \npr\ (of type {\tt integer}).
The selection of contributions to the Born amplitude remains 
unaffected.
The methods with $g_s$ in their name have the same effect and are
only kept for backward compatibility.
In fact, all methods (un-)select \underline{all} contributions.

\subsection{Updates for process computation}
\label{computation}

The methods \texttt{get\_*\_amplitude\_rcl} have been extended to support the
generalized treatment of fundamental types of couplings.
The following methods are concerned
\bei
\item {\tt get\_amplitude\_rcl (npr,pow,order,colour,hel,A)}\\
\pow\ is an overloaded argument and can be either an
 {\tt integer} specifying the power in $g_s$, or an array of {\tt integer}
 values specifying general powers in types of couplings: \\
\begin{insertion}
\begin{tabular}{@{}ll@{}}
  \texttt{pow}=n:                & selects the contribution $g_s^n$ to the amplitude;\\
  \texttt{pow=[n,m,$\ldots$]}:   & selects the contribution $[g_s^n, e^m, \ldots]$ to the \\
                                 & amplitude.
\end{tabular}
\end{insertion}

\item
 {\tt get\_squared\_amplitude\_rcl (npr,pow,order,A2)},\\ 
 {\tt get\_polarized\_squared\_amplitude\_rcl (npr,pow,order,hel,A2h)},\\
 {\tt get\_colour\_correlation\_rcl (npr,pow,i1,i2,A2cc)},\\
 {\tt get\_spin\_correlation\_rcl (npr,pow,A2sc)}, \\
 {\tt get\_spin\_colour\_correlation\_rcl}
 {\tt (npr,pow,i1,i2,A2scc)} \\
\pow\ is an overloaded argument and can be either an
 {\tt integer} specifying the power in $\alpha_s$, or an array of {\tt integer}
 values, specifying general powers in  in types of couplings:\\
\begin{insertion}
\begin{tabular}{@{}ll@{}}
  \texttt{pow}=n:                & selects the contribution $\alpha_s^n$ to the
  amplitude\\
                                 & squared;\\
  \texttt{pow=[n,m,$\ldots$]}:   & selects the contribution $[g_s^n,
  e^m, \ldots]$ to the\\ 
                                 &  amplitude squared.
\end{tabular}
\end{insertion}

\eei 

\subsection{\Cpp interface}
\label{calling recola from cpp}
The \recolatwo{} package includes a \Cpp interface\footnote{This interface is
also used to link the original \recola{} version to {\sc Sherpa}.} with the same
naming conventions for the subroutines as given in Sections 4.2--4.6 in
\citere{Actis:2016mpe} and the new subroutines in Sections \ref{input exthiggs},
\ref{input subroutines}, and \ref{definition}.
The basic usage is identical to the one in \FortranNinety{} and follows the
computation flow presented at the beginning of \refse{calling recola}.  In order
to use \recolatwo{} in a \Cpp program the \recolatwo{} header file needs to be
included as follows:
\begin{verbatim}
  #include "recola.hpp"
\end{verbatim}
This gives access to the \recola{} namespace. The functions are called in the
typical \Cpp syntax, e.g.:
\begin{verbatim}
  Recola::define_process_rcl(1, "u u -> u u", "NLO");
\end{verbatim}
The namespace identifier ``\texttt{Recola::}'' can be omitted by importing
\recola{} as follows:
\begin{verbatim}
  #include "recola.hpp"
  using namespace Recola
\end{verbatim}
The \FortranNinety{} subroutines translate to functions in \Cpp with the
argument types being identified according to:
\begin{center}
  \begin{tabular}{|c|c|}
  \hline 
  \FortranNinety & \Cpp\\
  \hline 
  \hline 
    {\texttt integer} & {\texttt int} \\
    {\texttt integer, dimension(:)} & {\texttt int[\;]} \\
    {\texttt logical} & {\texttt bool} \\
    {\texttt real(dp)} & {\texttt double} \\
    {\texttt real(dp), dimension(:)} & {\texttt double[\;]} \\
    {\texttt complex(dp)} & {\texttt std::complex$\langle$double$\rangle$} \\
    {\texttt character(len=*)} & {\texttt std::string} \\
  \hline
\end{tabular}
\end{center}
For multi-dimensional arrays, such as the momenta {\tt p}, the conventions for
the order of the indices is the same as in \FortranNinety, and any necessary
transposition is performed within the interface. 
The complex numbers are the only additional \Cpp Standard Library dependency
used in the interface.

Return values are handled by call-by-reference, thus, all \Cpp functions are
declared as void functions.  Optional arguments are implemented via function
overloading, i.e.\ missing arguments are replaced by default values. For
instance, the call
\begin{verbatim}
  Recola::use_alphaZ_scheme_rcl (); 
\end{verbatim}
enables the use of the $\alpha(M_{\rm Z})$ scheme (see Section 4.2.2 in
\citere{Actis:2016mpe}),
using the default value for $\alpha(M_{\rm Z})$ which is hard-coded in \recola.
Providing an explicit argument via
\begin{verbatim}
  Recola::use_alphaZ_scheme_rcl (0.0078125);
\end{verbatim}
allows to use a different value for $\alpha(M_{\rm Z})$ than the default in the
running session.

The original \FortranNinety{} demo files are available as \Cpp demo files.  The
compilation of the \Cpp demo files follows the same steps as given in 
\refse{demo files}, \ie by either running 
\begin{verbatim}
   make <demofile>
\end{verbatim}
in the build folder or directly via the run script
\begin{verbatim}
   ./run <demofile>
\end{verbatim}
in the demo folder, with \texttt{<demofile>} taking the values
\texttt{cdemo0\_rcl}, \texttt{cdemo1\_rcl}, \texttt{cdemo2\_rcl}, \texttt{cdemo3\_rcl}, \texttt{cdemo4\_rcl}, or
\texttt{cdemo5\_rcl}. The content of each \texttt{cdemo} file is identical to
the content of the corresponding (\FortranNinety) \texttt{demo} file.

A few \Cpp functions differ from the usage and naming convention of the
\FortranNinety{} subroutines owing to the conceptual difference of optional arguments
and arrays in  \FortranNinety{} and \Cpp.  Thus, the functions
\begin{itemize}
  \item {\tt use\_gfermi\_scheme\_rcl} (Section 4.2.18 in \citere{Actis:2016mpe}),
  \item {\tt set\_gs\_power\_rcl} (Section 4.3.2 in \citere{Actis:2016mpe}),
\end{itemize}
are replaced by the following ones:

\subsubsection{\tt use\_gfermi\_scheme\_rcl}
This \Cpp function takes no arguments and calls the subroutine {\tt
use\_gfermi\_scheme\_rcl(g,a)} (\FortranNinety) neither setting a value
for {\tt g} nor {\tt a}. This corresponds to selecting
the $G_{\rm F}$ scheme as renormalization scheme for the EW coupling and using
the default value for the Fermi constant $G_{\rm F}$ which is hard-coded in
\recola. 

\subsubsection{\tt use\_gfermi\_scheme\_and\_set\_gfermi\_rcl(g)}
This \Cpp function calls the subroutine {\tt
use\_gfermi\_scheme\_rcl(g,a)} (\FortranNinety), passing the value $g$ of type {\tt double}. 

\subsubsection{\tt use\_gfermi\_scheme\_and\_set\_alpha\_rcl(a)}
This \Cpp function calls the subroutine {\tt
use\_gfermi\_scheme\_rcl(g,a)} (\FortranNinety), passing the value $a$ of type {\tt double}.

\subsubsection{\tt set\_gs\_power\_rcl(npr,gsarray,gslen)}
This \Cpp function calls the subroutine {\tt set\_gs\_power\_rcl(npr, gsarray)}
(\FortranNinety), passing the value of {\tt npr} (of type {\tt int}) and {\tt
gsarray} (of type {\tt int[][2]}).  The value {\tt gslen} is the length of the
first index of {\tt gsarray} which needs to be passed explicitly to \FortranNinety.

\subsubsection{Missing subroutines}
The subroutines {\tt get\_colour\_configurations\_rcl} (Section 4.5.5 in
\citere{Actis:2016mpe}) and {\tt get\_helicity\_configurations\_rcl} (Section
4.5.6 in \citere{Actis:2016mpe}) are currently not included in the \Cpp
interface. The authors are willing to provide a custom solution, upon request,
to include their functionality.

\subsection{\Python{} interface}
\label{calling recola from python}
The \recolatwo{} package includes a \Python{} interface with the same naming conventions
for the subroutines as given in Sections 4.2--4.6 in \citere{Actis:2016mpe} and
the new subroutines in Sections \ref{input exthiggs}, \ref{input subroutines},
and \ref{definition}.
The \Python{} interface requires to build an additional
library, called \texttt{pyrecola}, which is done automatically
alongside the \recolatwo{} library if the
\Python{} libraries are found on the system and if \recolatwo{} is build as a
shared library.
See \refse{demo files} for more details on building \texttt{pyrecola}.
The basic usage is identical to the one in \FortranNinety{} and follows the
computation flow presented at the beginning of \refse{calling recola}.  In order
to use \recolatwo{} in a \Python{} program the \recolatwo{} library needs to be
loaded as follows:
\begin{verbatim}
  import pyrecola
\end{verbatim}
This step requires the python library \texttt{pyrecola.so} to be present in the current working
directory, or, alternatively, by updating the environment path
\texttt{PYTHONPATH} as follows
\begin{verbatim}
  export PYTHONPATH=$PYTHONPATH:<PATH_TO_PYRECOLA>
\end{verbatim}
where \texttt{<PATH\_TO\_PYRECOLA>} is the absolute path to the directory
containing
\texttt{pyrecola.so}.
Once the \recolatwo{} namespace is accessible the functions are called in the
typical \Python{} syntax, e.g.:
\begin{verbatim}
  pyrecola.define_process_rcl(1, 'u u -> u u', 'NLO')
\end{verbatim}
The namespace identifier {\frenchspacing``\texttt{pyrecola}.''} can be omitted by importing
\recolatwo{} as follows:
\begin{verbatim}
  from pyrecola import *
\end{verbatim}
Instead of loading all methods via \texttt{*} specific ones can be imported,
\eg{} 
\begin{verbatim}
  from pyrecola import (define_process_rcl,
                        generate_process_rcl,
                        compute_process_rcl)
\end{verbatim}
The \FortranNinety{} subroutines translate to functions in \Python{} with the
argument types being identified according to:
\begin{center}
  \begin{tabular}{|c|c|}
  \hline 
  \FortranNinety & \Python\\
  \hline 
  \hline 
    {\texttt integer} & {\texttt int} \\
    {\texttt integer, dimension(:)} & {\texttt list} \\
    {\texttt logical} & {\texttt bool} \\
    {\texttt real(dp)} & {\texttt float} \\
    {\texttt real(dp), dimension(:)} & {\texttt list} \\
    {\texttt complex(dp)} & {complex} \\
    {\texttt character(len=*)} & {str} \\
  \hline
\end{tabular}
\end{center}
For multi-dimensional arrays, such as the momenta {\tt p}, the conventions for
the order of the indices is the same as in \FortranNinety, and any necessary
transposition is performed within the interface.  Note that explicit type
checking is performed in the \Python/\CC{} API interface (\texttt{pyrecola.c}).

In contrast to the \FortranNinety{} and \Cpp{} interface,
computed results are returned by function calls.
For example, the call
\begin{Verbatim}[commandchars=\\\{\}]
  als = get_alphas_rcl()
\end{Verbatim}
returns the value of $\alpha_s$ and stores it in the variable \texttt{als}.
In general, the return value is a tuple of variables. 
The return type of a function can be inquired from the documentation as
described below or inferred from the \Python/\CC{} API interface \texttt{pyrecola.c}.

Furthermore, optional arguments are not implemented by operator
overloading, but via keyword arguments which allow most of the
\Python{} methods to be called in the same way as done in
\FortranNinety{} with a few exceptions discussed below.  For instance,
the call
\begin{verbatim}
  pyrecola.use_alphaZ_scheme_rcl()
\end{verbatim}
enables the use of the $\alpha(M_{\rm Z})$ scheme,
using the default value for $\alpha(M_{\rm Z})$ which is hard-coded in
\recolatwo.
Providing an explicit argument as
\begin{verbatim}
  pyrecola.use_alphaZ_scheme_rcl (a=0.0078125)
\end{verbatim}
allows to use a different value for $\alpha(M_{\rm Z})$ than the default in the
running session.

The \Python{} interface comes with a built-in documentation for every
public method in \recolatwo{} and can be accessed by calling
\texttt{help} inside \Python{} on either the module \texttt{pyrecola}
itself, or on specific methods.  For instance, calling
\texttt{help(use\_alphaZ\_scheme\_rcl)} returns

{\footnotesize
\begin{Verbatim}[commandchars=\\\{\}]
  >>> import pyrecola
  >>> help(pyrecola.use_alphaZ_scheme_rcl)
  Help on built-in function use_alphaZ_scheme_rcl in module pyrecola:

  {use_alphaZ_scheme_rcl(...)}
      use_alphaZ_scheme_rcl(a=None)
      
      Sets the EW renormalization scheme to the alphaZ scheme.
      a -> Sets value of alpha to `a`
      For `a=None` (DEFAULT) the hard-coded value for alpha is used.
\end{Verbatim}
}

The original \FortranNinety{} demo files are available as \Python{} demo files.
No compilation of the demo files is required and they can be run
 directly by executing 
\begin{verbatim}
   python <demofile>
\end{verbatim}
in the demo folder, with \texttt{<demofile>} taking the values
\texttt{pydemo0\_rcl.py}, \texttt{pydemo1\_rcl.py}, \texttt{pydemo2\_rcl.py}, 
\texttt{pydemo3\_rcl.py}, \texttt{pydemo4\_rcl.py} or \texttt{pydemo5\_rcl.py}.
The content of each \texttt{pydemo} file is identical to
the content of the corresponding (\FortranNinety) \texttt{demo} file.

A few \Python{} functions differ from the usage of the
\FortranNinety{} subroutines owing to the conceptual difference of optional
arguments in \FortranNinety{} and \Python.  The following functions are
concerned
\begin{itemize}
  \item {\tt get\_amplitude\_rcl\\(npr,order,colour,hel,pow=None,gs=None)}
  \item {\tt get\_squared\_amplitude\_rcl\\(npr,order,pow=None,als=None)}
  \item {\tt
    get\_polarized\_squared\_amplitude\_rcl\\(npr,order,hel,pow=None,als=None)}
  \item {\tt get\_colour\_correlation\_rcl\\(npr,i1,i2,pow=None,als=None)}
  \item {\tt get\_spin\_correlation\_rcl\\(npr,pow=None,als=None)}
  \item {\tt get\_spin\_colour\_correlation\_rcl\\(npr,i1,i2,pow=None,als=None)}
\end{itemize}
For all of these functions the difference is due to the new treatment of powers
in \recolatwo{} and the backward compatibility with \recola, as described in \refse{definition}.
The argument \texttt{pow}, which is overloaded in the \FortranNinety{}
interface, is replaced in \Python{} by the keyword arguments \texttt{pow} and
\texttt{gs} (or \texttt{als}).\footnote{We refer to the subroutines in Section (4.5) in
\citere{Actis:2016mpe} for the description of the other arguments.}
  The functions are called in the usual way with
the exception that either \texttt{pow} or \texttt{gs}/\texttt{als} is
passed and labelled explicitly, e.g.:
\begin{verbatim}
  A2 = pyrecola.get_squared_amplitude_rcl(..., pow=[2,4])
  A2 = pyrecola.get_squared_amplitude_rcl(..., als=1)
\end{verbatim}

\subsubsection{Missing subroutines}
The subroutines {\tt get\_colour\_configurations\_rcl} (Section 4.5.5 in
\citere{Actis:2016mpe}) and {\tt get\_helicity\_configurations\_rcl} (Section
4.5.6 in \citere{Actis:2016mpe}) are currently not included in the \Python{}
interface.
The authors are willing to provide a custom solution, upon request, to include
their functionality.

\section{Conclusions}
\label{conclusions}

The \recolatwo{} library computes amplitudes in the Standard Model of
particle physics including QCD and electroweak interaction and in
general quantum field theories at the tree and one-loop level with no
a-priori restriction on the particle multiplicities, once
corresponding model files are available. Amplitudes can be obtained
for specific colour structures and helicities and squared amplitudes
with or without summation/average over helicities. We provide
subroutines for the computation of colour- and spin-correlated
leading-order squared amplitudes that are required in the dipole
subtraction formalism. Furthermore, the code supports the selection of
resonant contributions allowing for the computation of factorizable
corrections in pole approximations.

In this first release of \recolatwo{} we include \recolatwo{} model
files for the computation of processes in the Two-Higgs-Doublet Model
and the Higgs-singlet extension of the Standard Model. The model files
are generated in the complex-mass scheme, and various renormalization
schemes are supported for the electromagnetic coupling, the strong
coupling and additional parameters in extended Higgs sectors.

The present version of \recolatwo{} is restricted to theories with
scalars, Dirac fermions and vector bosons. An enhanced version with
Majorana-fermion support and further \recolatwo{} model files,
including anomalous couplings, is in preparation.

\section{Acknowledgements}
We thank B.~Biedermann, M.~Chiesa, R.~Feger and M.~Pellen for
performing various checks of the code.  A.~D. and J.-N.~L. acknowledge
support from the German Research Foundation (DFG) via grants DE
623/4-1 and DE 623/5-1. The work of J.-N.~L. is supported by the
Studienstiftung des Deutschen Volkes. The work of S.~U. was supported
in part by the European Commission through the ``HiggsTools'' Initial
Training Network PITN-GA-2012-316704.
The research of A.D. and S.U. was supported in part by the Munich Institute for
Astro- and Particle Physics (MIAPP) of the DFG cluster of excellence ``Origin and
Structure of the Universe''.

\appendix

\section{Checks}
\label{appendix_checks}
\recolatwo{} has been thoroughly tested against \recola, guaranteeing
the consistency for all checks of Appendix~B in
\citere{Actis:2016mpe} for the SM.

In the \THDM{} all renormalized scalar two-point functions of the
extended Higgs sector in the \THDM{} have been verified off-shell
against an independent approach in \QGraf{} \cite{Nogueira:1991ex} and
\QGS, which is an extension of \GraphShot{} \cite{Actis}.
Furthermore, we have compared all partonic channels to Higgs decays
into four fermions against the independent calculation{}
\cite{Altenkamp:2017ldc} based on \FeynArts/\FormCalc{}
\cite{Hahn:2000kx,Gross:2014ola}.  Finally, all models have been
tested against the BFM implementation \cite{Denner:2017vms}.

\bibliographystyle{elsarticle-num}
\bibliography{recola2}

\end{document}